\documentclass[journal]{IEEEtran}
\usepackage{times,amsmath,amssymb,amsfonts,epsfig,graphicx}
\usepackage{enumerate}
\usepackage{stackrel}
\usepackage{multirow}
\usepackage{empheq}
\usepackage{color}
\usepackage{booktabs}
\ifCLASSOPTIONcompsoc
\usepackage[tight,normalsize,sf,SF]{subfigure}
\else
\usepackage[tight,footnotesize]{subfigure}
\fi

\newcommand{\vectornorm}[1]{\|#1\|}

\DeclareMathOperator*{\argmax}{arg\,max}
\newtheorem{prop}{Proposition}
\newtheorem{corollary}{Corollary}
\newcommand{\eqdef}{\stackrel{\triangle}{=}}
\begin{document}%
\title{Fixed-rank Rayleigh Quotient Maximization\\by an $M$PSK Sequence}%
\author{Anastasios~Kyrillidis,~\IEEEmembership{Student~Member,~IEEE,}
        and~George~N.~Karystinos,~\IEEEmembership{Member,~IEEE}%
\thanks{Manuscript received June 13, 2013; revised November 30, 2013; accepted January 16, 2014. The associate editor coordinating the review of this paper and approving it for publication was M. Juntti.}%
\thanks{This work was supported by the European Union (European Social Fund - ESF) and Greek national funds through the Operational Program ``Education and Lifelong Learning'' of the National Strategic Reference Framework (NSRF) Research Funding Program ``Thales - Investing in knowledge society through the European Social Fund.''
This paper was presented in part at the IEEE International Conference on Acoustics, Speech, and Signal Processing (ICASSP 2011), Prague, Czech Republic, May 2011.}%
\thanks{A. Kyrillidis is with the School of Computer \& Communication Sciences, EPFL, CH-1015 Lausanne, Switzerland (e-mail: anastasios.kyrillidis@epfl.ch).}%
\thanks{G. N. Karystinos is with the Department of Electronic and Computer Engineering, Technical University of Crete, Chania, 73100, Greece (e-mail: karystinos@telecom.tuc.gr).}}%

\markboth{This paper is to appear in IEEE Transactions on Communications}%
{Shell \MakeLowercase{\textit{et al.}}: Bare Demo of IEEEtran.cls for Journals}%
\maketitle%

\begin{abstract}%
Certain optimization problems in communication systems, such as limited-feedback constant-envelope beamforming or noncoherent $M$-ary phase-shift keying ($M$PSK) sequence detection, result in the maximization of a fixed-rank positive semidefinite quadratic form over the $M$PSK alphabet.
This form is a special case of the Rayleigh quotient of a matrix and, in general, its maximization by an $M$PSK sequence is $\mathcal{NP}$-hard.
However, if the rank of the matrix is not a function of its size, then the optimal solution can be computed with polynomial complexity in the matrix size.
In this work, we develop a new technique to efficiently solve this problem by utilizing auxiliary continuous-valued angles and partitioning the resulting continuous space of solutions into a polynomial-size set of regions, each of which corresponds to a distinct $M$PSK sequence.
The sequence that maximizes the Rayleigh quotient is shown to belong to this polynomial-size set of sequences, thus efficiently reducing the size of the feasible set from exponential to polynomial.
Based on this analysis, we also develop an algorithm that constructs this set in polynomial time and show that it is fully parallelizable, memory efficient, and rank scalable.
The proposed algorithm compares favorably with other solvers for this problem that have appeared recently in the literature.
\end{abstract}%
\begin{IEEEkeywords}%
Algorithms,
maximum likelihood detection,
MIMO systems,
noncoherent communication,
optimization methods,
phase shift keying,
Rayleigh quotient,
sequences.%
\end{IEEEkeywords}%

\IEEEpeerreviewmaketitle

\section{Problem Statement, Prior Work,\\and Contribution}
\label{sec:intro}
\subsection{Problem statement}

We consider the optimization problem
\begin{empheq}[box=\fbox]{align}
{\mathcal P}: \;\;\;\mathbf{s}_{\text{opt}} \eqdef \argmax_{\mathbf{s} \in \mathcal{A}_M^N} \vectornorm{\mathbf{V}^\mathcal{H} \mathbf{s}}
\label{eq:P}
\end{empheq}
where $ \mathcal{A}_M \eqdef \left\{ e^{\frac{j2\pi m}{M}} \; \big | \; m = 0,1, \dots , M-1 \right\} $ is the $M$-ary phase-shift keying ($M$PSK) alphabet, $\mathbf{V} \in \mathbb{C}^{N \times D}$ is a full-rank matrix, $\cdot^{\mathcal{H}}$ denotes Hermitian transpose operation, and $\|\cdot\|$ is the Euclidean $\ell_2$-norm.
Problem $\mathcal{P}$ in~(\ref{eq:P}) can be recast as
the special case of
the maximization of the Rayleigh quotient of $\mathbf{V}\mathbf{V}^\mathcal{H}$ over the $\mathit{M}$PSK alphabet, i.e.,
\begin{equation}
\mathbf{s}_\text{opt}=\argmax_{\mathbf{s}\in\mathcal{A}^N_M}\frac{\mathbf{s}^\mathcal{H}\mathbf{V}\mathbf{V}^\mathcal{H}\mathbf{s}}{\mathbf{s}^\mathcal{H}\mathbf{s}},
\label{eq:Rayleigh}
\end{equation}
and solved by an exponential-complexity exhaustive search among $M^{N-1}$ length-$N$ sequences.\footnote{The first element of $\mathbf{s}$ can be arbitrarily set to $1$ without losing optimality in~(\ref{eq:P}) or~(\ref{eq:Rayleigh}).}
However, such a solver would be impractical even for moderate values of the problem size $N$.

Of particular interest is the case where $\mathbf{V}$ in $\mathcal{P}$ is full-column-rank (i.e., it is a ``tall'' matrix) and its rank $D$ is independent of its row dimension $N$, which appears in certain optimization problems in communication systems, such as limited-feedback multiple-input multiple-output (MIMO) beamforming~\cite{Heath1998}-\cite{Santipach2011} and noncoherent sequence detection~\cite{MLSD:00}-\cite{STBC}.
For this particular fixed-$D$ case, $\mathcal{P}$ is no longer $ \mathcal{NP} $-hard if $\mathbf{s}$ belongs to the 2PSK or 4PSK alphabet~\cite{CG:00},~\cite{RankD:00},~\cite{CDMA:01} or general $M$PSK alphabet~\cite{Iran},~\cite{MLSD:00}; i.e., if $D$ is independent of $N$, then solving $\mathcal{P}$ requires only polynomial complexity in $N$.
The underlying principle of all the above works is the construction of a feasible set (i.e., a set of candidate sequences that contains the solution of $\mathcal{P}$) $\mathcal{S}(\mathbf{V})\subset\mathcal{A}^N_M$ which has polynomial cardinality $\left|\mathcal{S}(\mathbf{V})\right|$ and can be built with polynomial complexity.
After $\mathcal{S}(\mathbf{V})$ is constructed, the optimal sequence $\mathbf{s}_\text{opt}$ can be identified by a polynomial-complexity exhaustive search among the elements of $\mathcal{S}(\mathbf{V})$.

In this work, we present a new algorithm to solve $\mathcal{P}$ for any even%
\footnote{Although our algorithm can treat any (odd or even) $M\geq1$, to simplify the presentation we consider $M$ to be even.}
$M$ and arbitrary $D$ and prove that it has lower complexity than the current
state of the art~\cite{Iran},~\cite{MLSD:00},~\cite{CG:00},
is fully parallelizable and rank-scalable, and requires minimum memory resources.

\subsection{Prior work}

\noindent%
\underline{Case {\it(i)}: $M = 2$.}
A lot of effort has been made to solve $\mathcal{P}$ when $M=2$, i.e., $\mathbf{s}$ is a binary%
\footnote{In this work, a sequence is called binary if and only if each element of it equals $+1$ or $-1$.
In contrast, if each element of it equals $0$ or $1$, then the sequence is said to belong to the $ 0/1 $ alphabet.}
sequence, and $\mathbf{V}$ is a real-valued matrix.
We note that, for the special case $D=1$, ${\bf V}$ becomes a $N\times1$ vector and the solution of $\mathcal{P}$ is simply $\mathbf{s}_\text{opt}=\text{sgn}(\mathbf{V})$.%
\footnote{For any vector ${\bf v}\in{\mathbb R}^N$, we denote by $\text{sgn}(\mathbf{v})$ the vector ${\bf x}\in\left\{\pm1\right\}^N$ such that $x_n=1$ if $s_n>0$ and $x_n=-1$ if $s_n\leq0$, $n=1,2,\ldots,N$.}
Equivalently, we can say that $\mathcal{S}(\mathbf{V})=\left\{\text{sgn}(\mathbf{V})\right\}$ which has cardinality $1$ and is constructed with complexity $\mathcal{O}(N)$.
This simple case provides evidence that $\mathcal{P}$ may be polynomially solvable when the rank of $\mathbf{V}$ is independent of its row dimension.

The (more interesting) case of $D>1$ was considered in~\cite{CG:00},~\cite{CDMA:01} where it was shown that $\mathcal{P}$ is equivalent to the maximization of a rank-$D$ quadratic form over the $0/1$ alphabet. This case has been proven \cite{Zon:00} to be polynomially solvable through a variety of computational-geometry algorithms, such as the incremental algorithm for cell enumeration in arrangements~\cite{Inc:00},~\cite{Edelsbrunner-book} and the reverse search~\cite{Rev:00},~\cite{Zon:01}.
Although the incremental algorithm in~\cite{Inc:00},~\cite{Edelsbrunner-book} is time-efficient with overall complexity $\mathcal{O}\left(N^{D-1}\right)$ to build the feasible set $\mathcal{S}\left({\bf V}\right)$ of size $\left|\mathcal{S}\left({\bf V}\right)\right|=\mathcal{O}\left(N^{D-1}\right)$, it becomes impractical even for moderate values of $D$, since it follows an ``incremental'' strategy to construct the feasible set $\mathcal{S}(\mathbf{V})$: it solves the problem inductively and, thus, is too complicated to be implemented.
Furthermore, the critical disadvantage of the incremental algorithm is its memory inefficiency, since it needs to store all the extreme points, all faces, and their incidences.
On the other hand, the highly parallelizable reverse search~\cite{Rev:00},~\cite{Zon:01} is memory efficient and constructs a set of $\left|\mathcal{S}(\mathbf{V})\right|=\sum_{d=0}^{D-1}\binom{N-1}{d}$ candidate sequences in ${\mathcal O}(N^D\textbf{LP}(N,D))$ time-complexity where $\textbf{LP}(N,D)$ denotes the time to solve a linear programming problem with $N$ inequalities in $D$ variables.
The work in~\cite{Megiddo} showed that $\textbf{LP}(N,D)=\mathcal{O}\left(N\right)$ in fixed dimensions, implying that the overall complexity of the reverse search to build $\mathcal{S}\left({\bf V}\right)$ is $\mathcal{O}\left(N^{D+1}\right)$.
We note, however, that, until today, the reverse search has not been extended to general $M$PSK alphabets.

From a different perspective, based on the auxiliary-angle approach that was originally introduced in~\cite{Mack:00} to solve $\mathcal{P}$ with complexity $\mathcal{O}(N\log N)$ when $D=1$ and $M\geq2$, efficient solutions of $\mathcal{P}$ for the binary-$\mathbf{s}$ ($M=2$), real-$\mathbf{V}$ case are presented in~\cite{rank2} for $D=2$, in~\cite{rank3} for $D=3$, and in~\cite{RankD:00} for $D>3$ with complexity $\mathcal{O}(N\log N)$, $\mathcal{O}(N^2\log N)$, and $\mathcal{O}(N^D)$, respectively.
The methodology utilizes $D-1$ auxiliary angles and partitions the ($D-1$)-dimensional hypercube into a polynomial-size set of distinct regions so that each region is associated with a distinct binary sequence.
The set of binary sequences that are obtained has the same size $\left|\mathcal{S}(\mathbf{V})\right|$ as the one produced by the reverse search~\cite{Rev:00},~\cite{Zon:01}.
However, the auxiliary-angle method is fully parallelizable and rank-scalable, requires minimum memory resources, and constructs the candidate solution set $\mathcal{S}(\mathbf{V})$ with lower complexity than the reverse search for any $D\geq2$.

\noindent%
\underline{Case {\it(ii)}: $M = 4$.}
The reverse-search~\cite{Rev:00},~\cite{Zon:01} and auxiliary-angle~\cite{RankD:00},~\cite{rank2},~\cite{rank3} methods were originally considered as potential solvers of $\mathcal{P}$ for $M=2$ and real $\mathbf{V}$.
If $\mathbf{s}$ is binary but $\mathbf{V}$ is complex, then $\mathcal{P}$ can be rewritten with $\mathbf{V}$ substituted by a real matrix of size $N\times2D$.
Hence, the above methods can still solve $\mathcal{P}$ with complexity that is polynomial in $N$ and determined by $2D$ (instead of $D$).
Finally, if $\mathbf{s}$ is quaternary (i.e., $M=4$), then $\mathcal{P}$ can be rewritten with $\mathbf{s}$ and $\mathbf{V}$ substituted by a binary sequence of length $2N$ and a real matrix of size $2N\times2D$, respectively, and solved by the reverse-search or auxiliary-angle methods with complexity that is still polynomial and determined by $2N$ and $2D$ (instead of $N$ and $D$).
Due to their ease of implementation, both techniques have been used for maximum-likelihood (ML) noncoherent detection of uncoded~\cite{CG:00} or space-time coded sequences of 2PSK or 4PSK signals~\cite{STBC} and near-ML multiuser detection~\cite{CDMA:01}.

\noindent%
\underline{Case {\it(iii)}: Arbitrary $M$.}
If $M>4$, then $\mathcal{P}$ is still polynomially solvable, as shown in~\cite{Mack:00} for $D=1$ and~\cite{Iran},~\cite{MLSD:00} for any $D\geq1$.
Specifically, in the context of ML noncoherent detection of
$M$PSK
in single-antenna systems,~\cite{Mack:00} presented a method (which reappeared later in~\cite{sweldens}) that solves $\mathcal{P}$ when $D=1$ by constructing a candidate solution set of size $\left|\mathcal{S}(\mathbf{V})\right|=N$ with complexity $\mathcal{O}(N\log N)$.
In~\cite{MLSD:00}, it was proven that $\mathcal{P}$ is solvable in polynomial time for any $M\geq1$ and $D\geq1$ by constructing a candidate sequence set of size $\left|\mathcal{S}(\mathbf{V})\right|=\mathcal{O}\left(N^{2D}\right)$ with complexity and storage requirement $\mathcal{O}\left(N^{2D}\right)$, based on the incremental algorithm in~\cite{Inc:00},~\cite{Edelsbrunner-book}.
However, as mentioned above, the incremental algorithm is of purely theoretical value.
Therefore, although~\cite{MLSD:00} was the first work that identified the potential of polynomial solvability of $\mathcal{P}$ for this configuration, it did not offer a practical algorithm to solve it.
In~\cite{SIMO:00}, the authors built on the auxiliary-angle methodology of~\cite{RankD:00},~\cite{Mack:00}-\cite{sweldens} to develop a practical algorithm that constructs $\mathcal{S}\left({\bf V}\right)$ with complexity
$\mathcal{O}\left(N^{2D}\right)$.
Although it is proven that $\left|\mathcal{S}\left({\bf V}\right)\right|=\mathcal{O}\left(N^{2D-1}\right)$, the algorithm in~\cite{SIMO:00} is not optimal with respect to the actual size of $\mathcal{S}\left({\bf V}\right)$, since, by construction, it produces multiple phase-rotated candidate sequences that are equivalent with respect to the optimization metric in $\mathcal{P}$.
Finally, in the context of limited-feedback MIMO beamforming,~\cite{Iran}  presented a Voronoi-cell based algorithm that, for any $M\geq1$ and $D\geq1$, builds a candidate sequence set of size $\left|\mathcal{S}(\mathbf{V})\right|=\mathcal{O}\left(N^{2D-1}\right)$ with complexity $\mathcal{O}\left(N^{2D}\right)$.

\subsection{Contribution}
\label{subsec:contribution}

\begin{table*}[t!]
\renewcommand{\arraystretch}{1.7}
\caption{Complexity Comparison}
\label{table:complexity}
\centerline{
\begin{tabular}{|c|c|c|}
\toprule
Method & Size of Candidate Sequence Set $\mathcal{S}(\mathbf{V})$ & Time-complexity\\
\midrule \midrule
\cite{Iran}&$\binom{MN}{2D-1}\leq\left|\mathcal{S}\left({\bf V}\right)\right|=\mathcal{O}\left(N^{2D-1}\right)$&$\mathcal{O}\left(N^{2D}\right)$\\
\midrule
\cite{MLSD:00}&$\left|\mathcal{S}\left({\bf V}\right)\right|=\mathcal{O}\left(N^{2D}\right)$&$\mathcal{O}\left(N^{2D}\right)$\\
\midrule
\cite{SIMO:00}&$\left|\mathcal{S}\left({\bf V}\right)\right|=\mathcal{O}\left(N^{2D-1}\right)$&$\mathcal{O}\left(N^{2D}\right)$\\
\midrule
\cite{CG:00}&$\begin{array}{c}\left|\mathcal{S}\left({\bf V}\right)\right|=\sum_{d=0}^{2D-1}\binom{N-1}{d}=\mathcal{O}\left(N^{2D-1}\right),\;\;\text{if}\;M=2\\\left|\mathcal{S}\left({\bf V}\right)\right|=\sum_{d=0}^{2D-1}\binom{2N-1}{d}=\mathcal{O}\left(N^{2D-1}\right),\;\;\text{if}\;M=4\end{array}$&$\mathcal{O}\left(N^{2D+1}\right)$\\
\midrule
Proposed&$\left|\mathcal{S}\left({\bf V}\right)\right|=\sum_{d=1}^{D}\sum_{i=0}^{d-1}\binom{N}{i}\binom{N-i}{2(d-i)-1}\Big(\frac{M}{2}\Big)^{2(d-i)-2}\Big(\frac{M}{2}-1\Big)^{i}=\mathcal{O}\left(N^{2D-1}\right)$&$\mathcal{O}\left(N^{2D}\right)$\\
\bottomrule
\end{tabular}}
\end{table*}

In the present work, we follow the principles of the work in~\cite{RankD:00} to treat the complex-domain problem $\mathcal{P}$ for any even $M\geq2$ and any $D\geq1$.
Specifically, we introduce $2D-1$ auxiliary continuous angles and partition the ($2D-1$)-dimensional hypercube into a polynomial-size set of distinct regions, each of which is associated with a distinct $M$PSK sequence of length $N$.
The proposed algorithm is based on the framework presented in~\cite{SIMO:00} but confronts the problem in a more solid and optimized manner.
In particular, we manage to remove equivalent $M$-plicate candidates from $\mathcal{S}(\mathbf{V})$ that attain the same metric value in the objective function in $\mathcal{P}$ and are present in the solution proposed in~\cite{SIMO:00}, thus decreasing the size of $\mathcal{S}(\mathbf{V})$.
This way, we succeed to prove that the size of the candidate set for given $M \geq 2,~D \geq 1$ and $N$ is exactly given by
\begin{align}
&\left|\mathcal{S}(\mathbf{V})\right|\label{eq:SV}\\
&=\sum_{d = 1}^{D} \sum_{i = 0}^{d-1} \binom{N}{i}\binom{N - i}{2(d-i) - 1}\Big(\frac{M}{2}\Big)^{2(d-i)-2}\Big(\frac{M}{2} - 1\Big)^{i}.\;\;\;\;\;\;\nonumber
\end{align}
We emphasize that the proposed algorithm constructs the candidate set $\mathcal{S}(\mathbf{V})$ with complexity $\mathcal{O}\left(N^{2D}\right)$ and claim that the size of the produced candidate set provided in~(\ref{eq:SV}) is the smallest size that can be obtained, as of today, as compared to the present
state of the art~\cite{Iran},~\cite{MLSD:00},~\cite{CG:00}.
In addition, we show that the proposed algorithm is fully parallelizable and rank-scalable and requires minimum memory resources.

In comparison with the prior work that we presented in the previous subsection, a few conclusions can be drawn from Table~\ref{table:complexity}.
We note that the method in~\cite{CG:00} (which uses the reverse search~\cite{Rev:00},~\cite{Zon:01}) is limited to $M=2$ and $M=4$ only.
For $M=2$, it computes $\sum_{d=0}^{2D-1}\binom{N-1}{d}$ candidates (that is, as many as the proposed algorithm), while, for $M=4$, it computes $\sum_{d=0}^{2D-1}\binom{2N-1}{d}$ candidates (that is, twice as many as the proposed algorithm). In both cases, the complexity is $\mathcal{O}\left(N^{2D+1}\right)$, i.e., one order of magnitude higher than the complexity of the proposed algorithm.
The work in~\cite{MLSD:00} only calculates the order of the size of the candidate set as $\left|\mathcal{S}(\mathbf{V})\right|=\mathcal{O}\left(N^{2D}\right)$ and the corresponding complexity to build $\mathcal{S}(\mathbf{V})$ as $\mathcal{O}\left(N^{2D}\right)$, but does not identify the exact number of candidates in $\mathcal{S}(\mathbf{V})$.
Similarly, in~\cite{Iran}, the authors do not mention the exact size of the constructed candidate sequence set, but only prove that its order is $\mathcal{O}\left(N^{2D-1}\right)$ and it is constructed with complexity $\mathcal{O}\left(N^{2D}\right)$.
From the presented algorithm in~\cite{Iran}, it can be seen that at least $\binom{MN}{2D-1}$ subsets of indices are examined and each subset may or may not produce candidate sequence(s).
Hence, at least $\binom{MN}{2D-1}$ ``potential'' candidate sequences are examined, implying that $\binom{MN}{2D-1}$ is a lower bound on the complexity that~\cite{Iran} requires to build $\mathcal{S}(\mathbf{V})$.

\begin{figure}[t!]%
\centerline{\includegraphics[width=\columnwidth]{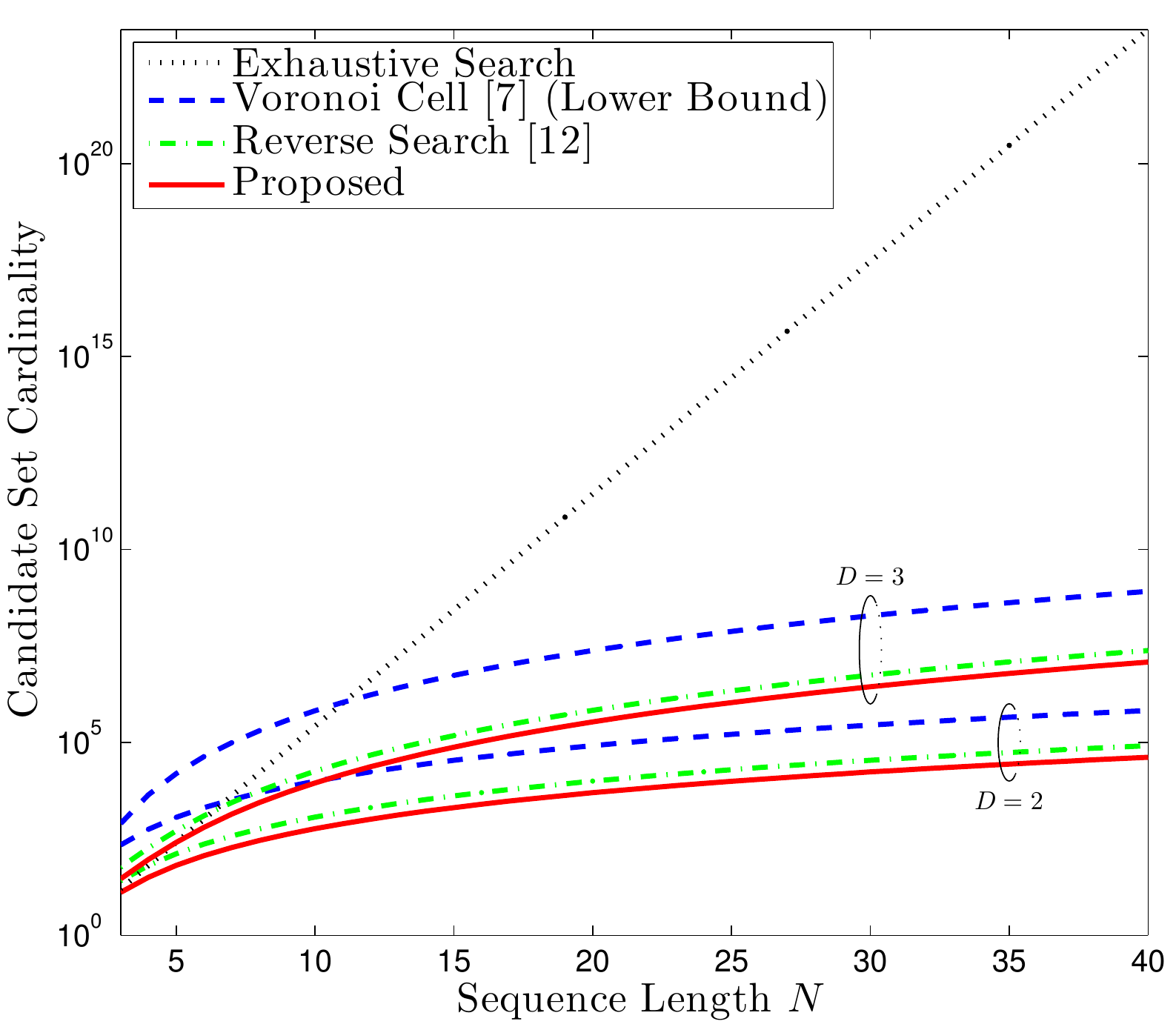}}
\caption{Cardinality of candidate sequence set $\mathcal{S}(\mathbf{V})$ for alphabet size $M=4$ and rank $D=2$ or $D=3$.
Actual values for the exhaustive search, the reverse search~\cite{CG:00}, and the proposed method and lower bound for the Voronoi-cell based approach~\cite{Iran}.}%
\label{fig:M4}%
\end{figure}

A comparison between the complexity of the current state of the art and the proposed algorithm is provided in Figs.~\ref{fig:M4} and~\ref{fig:M8}.
Since the complexity in Big-Oh notation of the works in~\cite{Iran},~\cite{MLSD:00}, and~\cite{SIMO:00} has the same order with the proposed algorithm and is lower than the complexity of the work in~\cite{CG:00}, the interest in comparing these algorithms is moving toward the exact size of the generated candidate sequence set.
The set that is generated by the method in~\cite{MLSD:00} is one order of magnitude larger than the one produced by the proposed algorithm, hence~\cite{MLSD:00} can be ignored in our comparison.
Regarding~\cite{SIMO:00}, as mentioned before, it produces a set that is larger than the proposed one by a factor of $M$, hence it can also be ignored.

As a result, in Fig.~\ref{fig:M4}, for $M=4$ and a fixed rank $D=2$ or $D=3$, we examine the size of the generated candidate sequence set $\mathcal{S}(\mathbf{V})$ as a function of the sequence length $N$ only for the Voronoi-cell based approach~\cite{Iran}, the reverse search~\cite{CG:00}, and the proposed algorithm.
For~\cite{Iran}, we plot the lower bound on $\left|\mathcal{S}(\mathbf{V})\right|$ from Table~\ref{table:complexity}.
For~\cite{CG:00}, we plot the exact size of $\mathcal{S}(\mathbf{V})$ for the case $M=4$, as shown in Table~\ref{table:complexity}.
For the proposed algorithm, we use the exact size of $\mathcal{S}(\mathbf{V})$ as provided in~(\ref{eq:SV}).
We observe that the proposed algorithm and the reverse search~\cite{CG:00} generate a set that is at least two orders of magnitude smaller than the lower-bound of the Voronoi-cell based approach~\cite{Iran}.
In fact, as can be seen in Fig.~\ref{fig:M4}, the reverse search generates a set that is twice as large as the one generated by the proposed algorithm.
In addition, as mentioned before, in comparison with the proposed algorithm, the reverse search requires one order of magnitude higher complexity to generate its candidate set.

Similar plots are presented in Fig.~\ref{fig:M8} for $M=8$ and a fixed rank $D=2$ or $D=3$.
We omit the reverse search, since it is not defined for $M>4$.
Once more, we plot the lower bound on $\left|\mathcal{S}(\mathbf{V})\right|$ from Table~\ref{table:complexity} for the Voronoi-cell based approach~\cite{Iran} and observe that our proposed algorithm generates at least two-three orders of magnitude less candidates.

Finally, it is interesting to examine the two limiting cases $D=1$ and $D=N$.
First, we observe that, if we set $D=1$ in~(\ref{eq:SV}), then we obtain $\left|\mathcal{S}(\mathbf{V})\right|=N$ which equals the size of the candidate sequence set of the algorithm in~\cite{Mack:00} (that, however, works only for $D=1$).
Therefore, in terms of the number of generated candidate sequences, our algorithm can also be seen as a generalization of~\cite{Mack:00} for any $D>1$.
In the other limiting case, if we set $D=N$ in~(\ref{eq:SV}), then, after a few calculations, we can show that $\left|\mathcal{S}(\mathbf{V})\right|=M^{N-1}$.
That is, if ${\bf V}$ is $N\times N$ and full-rank, then our algorithm generates all possible $M$-ary sequences of length $N$ that are not rotated versions of each other.
Of course, in such a case, one should directly operate a full-size exhaustive search without the need to build the candidate sequence set and our algorithm (as well as the state-of-the-art) would become meaningless.
Yet, the fact that for $D=N$ our algorithm constructively builds the $M^{N-1}$ sequences indicates that for lower ranks it may provide the minimum possible number of candidates that could be built with polynomial complexity; for the moment, the latter statement is only claimed as a conjecture.

The rest of the paper is organized as follows.
Section II is devoted to the theoretical developments of the proposed algorithm for solving $\mathcal{P}$ when $D$ is fixed.
The implementation of the proposed algorithm is discussed in more details in Section III.
Concluding remarks are drawn in Section IV.

{\it Notation:}
To simplify the presentation of our developments and proofs, we use a MATLAB-like notation.
Specifically, for any $i,j\in{\mathbb N}$ with $i\leq j$, we denote by $i:d:j$ the vector $[i\;\;i+d\;\;i+2d\;\ldots\;j]^T$; then, $i:j$ is a simplified notation for $i:1:j$.
For any $N\times1$ vector ${\bf x}$ and any ${\bf n}\in\{1,2,\ldots,N\}^K$, we denote by ${\bf x}_{\bf n}$ the $K\times1$ vector $\left[x_{n_1}\;x_{n_2}\;\ldots\;x_{n_K}\right]^T$.
Similarly, for any $N\times M$ matrix ${\bf X}$ and any ${\bf n}\in\{1,2,\ldots,N\}^K$, ${\bf m}\in\{1,2,\ldots,M\}^L$, we denote by ${\bf X}_{{\bf n},{\bf m}}$ the $K\times L$ matrix whose $(k,l)$-th element is ${\bf X}_{n_k,m_l}$;
then, ${\bf X}_{:,{\bf m}}$ and ${\bf X}_{{\bf n},:}$ are simplified notations for ${\bf X}_{1:N,{\bf m}}$ and ${\bf X}_{{\bf n},1:M}$, respectively.
Finally, $\Re\left\{{\bf X}\right\}$ and $\Im\left\{{\bf X}\right\}$ denote the real and imaginary parts of matrix ${\bf X}$.

\section{Efficient Fixed-rank Rayleigh Quotient Maximization by an $M$PSK Sequence}

\subsection{Problem Reformulation}

Without loss of generality (w.l.o.g.), we assume that each row of $\mathbf{V}$ in $\mathcal{P}$ has at least one nonzero element, i.e., $\mathbf{V}_{n,:} \neq \mathbf{0}$, $n=1,2,\ldots,N$.
If not, then the value of the variable $s_n$ related with the all-zero row of $\mathbf{V}$ would have no effect on the maximization procedure and could be simply ignored, reducing the dimension of our problem by $1$.

\begin{figure}[t!]%
\centerline{\includegraphics[width=\columnwidth]{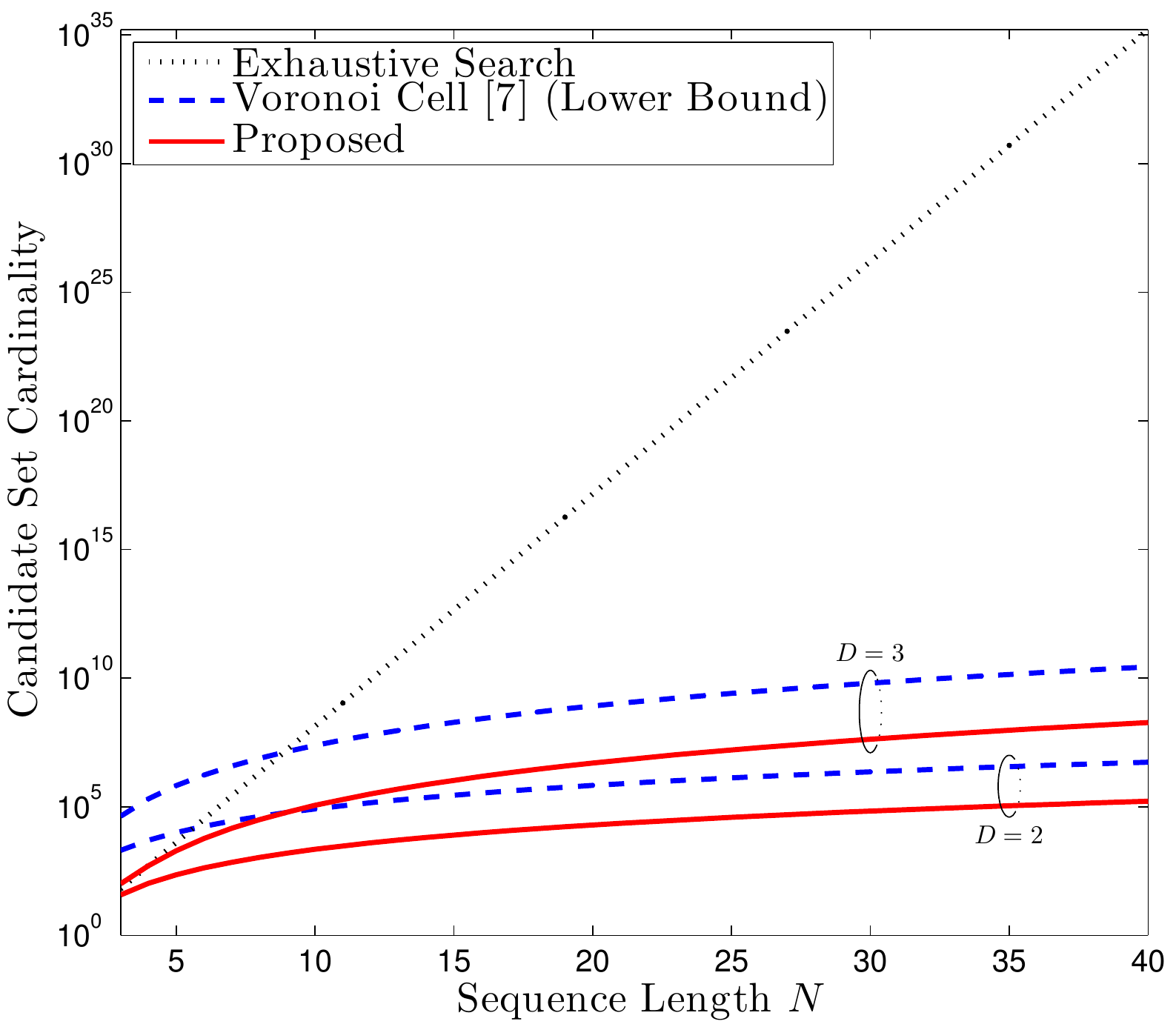}}
\caption{Cardinality of candidate sequence set $\mathcal{S}(\mathbf{V})$ for alphabet size $M=8$ and rank $D=2$ or $D=3$.
Actual values for the exhaustive search and the proposed method and lower bound for the Voronoi-cell based approach~\cite{Iran}.}%
\label{fig:M8}%
\end{figure}%

To develop an efficient method for the solution of $\mathcal{P}$ in~(\ref{eq:P}), we introduce the $(2D-1)\times1$ auxiliary-angle vector $\boldsymbol{\phi}\in(-\frac{\pi}{2},\frac{\pi}{2}]^{2D-2}\times(-\pi,\pi]$ and define the unit-norm $2D\times1$ real vector
\begin{equation}%
\tilde{\bf c}(\boldsymbol{\phi})\triangleq 
\begin{bmatrix}%
\sin\phi_1 \\
\cos\phi_1 \sin\phi_2 \\
\cos\phi_1 \cos\phi_2 \sin\phi_3 \\
\vdots \\
\left[\prod_{i=1}^{2D-2} \cos\phi_i \right]\sin\phi_{2D-1} \\
\left[\prod_{i=1}^{2D-2} \cos\phi_i \right]\cos\phi_{2D-1}
\end{bmatrix}%
\end{equation}%
and the unit-norm $D\times1$ complex vector%
\footnote{The definition of $\tilde{\bf c}(\boldsymbol{\phi})$ and ${\bf c}(\boldsymbol{\phi})$ differentiates the developments of this present work from the work in~\cite{SIMO:00} where the treatment of~(\ref{eq:P}) led to an algorithm of higher complexity by building on the auxiliary-angle methodology of~\cite{RankD:00},~\cite{Mack:00}-\cite{sweldens}.
In fact, the algorithm in~\cite{SIMO:00} is not optimal with respect to the actual size of the generated feasible set, since, by construction, it produces multiple phase-rotated candidate sequences that are equivalent with respect to the optimization metric in $\mathcal{P}$.
In this present work, the new definition of $\tilde{\bf c}(\boldsymbol{\phi})$ and ${\bf c}(\boldsymbol{\phi})$ helps us minimize the number of candidate sequences and overall complexity.}
\begin{align}%
\mathbf{c}(\boldsymbol{\phi})&\eqdef \tilde{\mathbf{c}}_{2:2:2D}(\boldsymbol{\phi}) + j \tilde{\mathbf{c}}_{1:2:2D-1}(\boldsymbol{\phi})
\label{eq_002}
\\
&=\begin{bmatrix}
\cos\phi_1 \sin\phi_2 + j \sin\phi_1 \\
\cos\phi_1 \cos\phi_2 \cos\phi_3 \sin\phi_4 + j \cos\phi_1 \cos\phi_2 \sin\phi_3 \\
\vdots \\
\left[\prod_{i=1}^{2D-1} \cos\phi_i \right] + j \left[\prod_{i=1}^{2D-2} \cos\phi_i \right] \sin\phi_{2D-1}
\end{bmatrix}.
\nonumber
\end{align}%
For notation simplicity, we set 
\begin{equation}%
\Phi\eqdef\left(-\frac{\pi}{2},\frac{\pi}{2}\right].
\end{equation}%

We observe that, as $\boldsymbol{\phi}$ varies in $\Phi^{2D-2}\times(-\pi,\pi]$, the real vector $\tilde{\bf c}(\boldsymbol{\phi})$ scans the entire unit-norm $2D$-dimensional real hypersphere.
At the same time, the complex vector ${\bf c}(\boldsymbol{\phi})$ scans the entire unit-norm $D$-dimensional complex hypersphere.
As we will see in the rest of this subsection, for any value of ${\bf c}(\boldsymbol{\phi})$ (that is, for any point of the hypersphere), our initial optimization problem is solved with linear complexity.
Therefore, one would need to compare against each other all solutions that are collected while the entire hypersphere is scanned.
Surprisingly, the number of solutions that are collected is polynomial in $N$.
This obervation is the motivation behind the definition and use of ${\bf c}(\boldsymbol{\phi})$ and will lead to the development of a polynomial-time algorithm in the following subsections.

From Cauchy-Schwarz Inequality, we observe that, for any $\mathbf{a}\in\mathbb{C}^D $,
\begin{equation}
\Re\left\{\mathbf{a}^\mathcal{H}\mathbf{c}(\boldsymbol{\phi})\right\}\leq\left|\mathbf{a}^\mathcal{H}\mathbf{c}(\boldsymbol{\phi})\right|\leq\left\|\mathbf{a}\right\|\underbrace{\left\|\mathbf{c}(\boldsymbol{\phi})\right\|}_{=1}=\left\|\mathbf{a}\right\|.
\label{eq_003}
\end{equation}
Equality is achieved in both inequalities in~(\ref{eq_003}), if and only if $ \boldsymbol{\phi} $ consists of the spherical coordinates of vector $ \mathbf{a} $, i.e., if and only if
\begin{equation}
\mathbf{c}(\boldsymbol{\phi})=\frac{\mathbf{a}}{\left\|\mathbf{a}\right\|},
\label{eq_004}
\end{equation}
since $\Re\left\{\mathbf{a}^\mathcal{H}\frac{\mathbf{a}}{\left\|\mathbf{a}\right\|}\right\}=\left\|\mathbf{a}\right\|$.
Using the above, our original problem $\mathcal{P}$ in~(\ref{eq:P}) is rewritten as
\begin{equation}
\max_{\mathbf{s}\in\mathcal{A}_M^N}\left\|\mathbf{V}^\mathcal{H}\mathbf{s}\right\|=\max_{\mathbf{s}\in\mathcal{A}_M^N}\max_{\boldsymbol{\phi}\in\Phi^{2D-2}\times(-\pi,\pi]}\Re\left\{\mathbf{s}^\mathcal{H}\mathbf{V}\mathbf{c}(\boldsymbol{\phi})\right\}.
\label{eq_005}
\end{equation}

Before we proceed, to reduce the overall complexity of our following developments, we further restrict the range of the last auxiliary angle $\phi_{2D-1}$ from $\left(-\pi,\pi\right]$ to $\left(-\frac{\pi}{M},\frac{\pi}{M}\right]$ without losing optimality in~(\ref{eq_005}).
The reason is that the phase-rotated sequences ${\bf s}$ and $e^{j\theta}{\bf s}$ result in the same value $\left\|{\bf V}^\mathcal{H}{\bf s}\right\|=\left\|{\bf V}^\mathcal{H}\left(e^{j\theta}{\bf s}\right)\right\|$ in~(\ref{eq_005}), for any $e^{j \theta} \in \mathcal{A}_M$.
Then, our problem in~(\ref{eq_005}) is rewritten as
\begin{equation}
\max_{\mathbf{s}\in\mathcal{A}_M^N}\left\|\mathbf{V}^\mathcal{H}\mathbf{s}\right\|=\max_{\mathbf{s}\in\mathcal{A}_M^N}\max_{\boldsymbol{\phi}\in\Phi^{2D-2}\times\left(-\frac{\pi}{M},\frac{\pi}{M}\right]}\Re\left\{\mathbf{s}^\mathcal{H}\mathbf{V}\mathbf{c}(\boldsymbol{\phi})\right\}.
\label{eq_009}
\end{equation}
A detailed proof of~(\ref{eq_009}) is provided in the Appendix.
By interchanging the maximizations in~(\ref{eq_009}), we obtain the equivalent problem
\begin{equation}
\max_{\mathbf{s}\in\mathcal{A}_M^N}\left\|\mathbf{V}^\mathcal{H}\mathbf{s}\right\|=\max_{\boldsymbol{\phi}\in \Phi^{2D-2} \times (-\frac{\pi}{M}, \frac{\pi}{M}]}\sum_{n=1}^N \max_{s_n\in\mathcal{A}_M}\Re\left\{s_n^{*}\mathbf{V}_{n,:}\mathbf{c}(\boldsymbol{\phi})\right\}.
\label{eq_010}
\end{equation}

\subsection{Candidate Sequence Set $\mathbf{\mathcal{S}}(\mathbf{V})$}

By the inner maximization rule in~(\ref{eq_010}), for any $D\times1$ complex vector $\mathbf{v}$, we define the $M$PSK decision function $d(\mathbf{v}^T; \boldsymbol{\phi})$ that maps $\boldsymbol{\phi}$ to $ \mathcal{A}_M $ according to
\begin{equation}
d(\mathbf{v}^T; \boldsymbol{\phi})\eqdef\argmax_{s\in\mathcal{A}_M}\Re\left\{s^*\mathbf{v}^T\mathbf{c}(\boldsymbol{\phi})\right\}
\label{eq_017}
\end{equation}
and, for any $N\times D$ complex matrix $\mathbf{V}$, we define, using~(\ref{eq_017}), the $M$PSK sequence decision function $\mathbf{d}\left({\bf V};\boldsymbol\phi\right)$ that maps $\boldsymbol{\phi}$ to ${\mathcal A}_M^N$ according to
\begin{equation}{\label{eq:decision}}
\mathbf{d}(\mathbf{V};\boldsymbol{\phi})\eqdef\argmax_{{\bf s}\in{\mathcal A}_M^N}\Re\left\{{\bf s}^{\mathcal H}{\bf V}\mathbf{c}(\boldsymbol{\phi})\right\}=%
\begin{bmatrix}
d(\mathbf{V}_{1,:}; \boldsymbol{\phi}) \\
d(\mathbf{V}_{2,:}; \boldsymbol{\phi}) \\
\vdots \\
d(\mathbf{V}_{N,:}; \boldsymbol{\phi}) \\
\end{bmatrix}.
\end{equation}
Computing $\mathbf{d}(\mathbf{V};\boldsymbol{\phi})$ for any $\boldsymbol{\phi}\in\Phi^{2D-2}\times\left(-\frac{\pi}{M},\frac{\pi}{M}\right]$, we collect all $M$PSK candidate sequences into set
\begin{equation}
\mathcal{S}(\mathbf{V})\eqdef\bigcup_{\boldsymbol{\phi}\in\Phi^{2D-2}\times\left(-\frac{\pi}{M},\frac{\pi}{M}\right]}\left\{\mathbf{d}\left(\mathbf{V};\boldsymbol{\phi}\right)\right\}\subseteq\mathcal{A}_M^N \label{eq_018}
\end{equation} 
and our problem $\mathcal{P}$ in~(\ref{eq_010}) becomes
\begin{equation}
\max_{\mathbf{s}\in\mathcal{A}_M^N}\left\|\mathbf{V}^\mathcal{H}\mathbf{s}\right\|=\max_{\mathbf{s}\in\mathcal{S}(\mathbf{V})}\left\|\mathbf{V}^\mathcal{H}\mathbf{s}\right\|,
\end{equation}
i.e., the $M$PSK candidate sequence $\mathbf{s}_{\text{opt}}$ that maximizes the metric of interest in~(\ref{eq:P}) belongs to $\mathcal{S}(\mathbf{V})$.

\subsection{Decision Boundaries}

\begin{figure*}[t!]
\centerline{\includegraphics[width=3.5in]{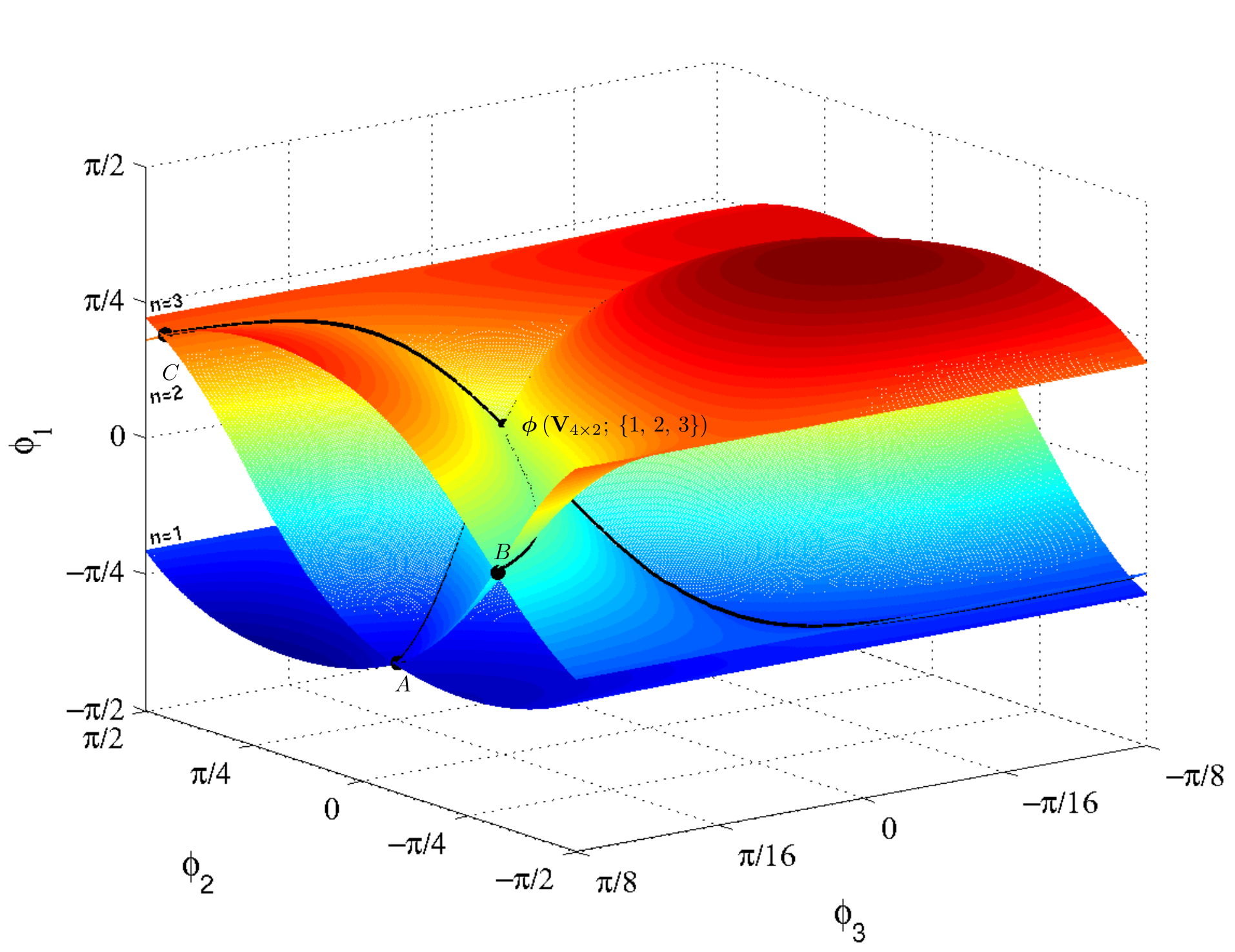}
\hfill
\includegraphics[width=3.5in]{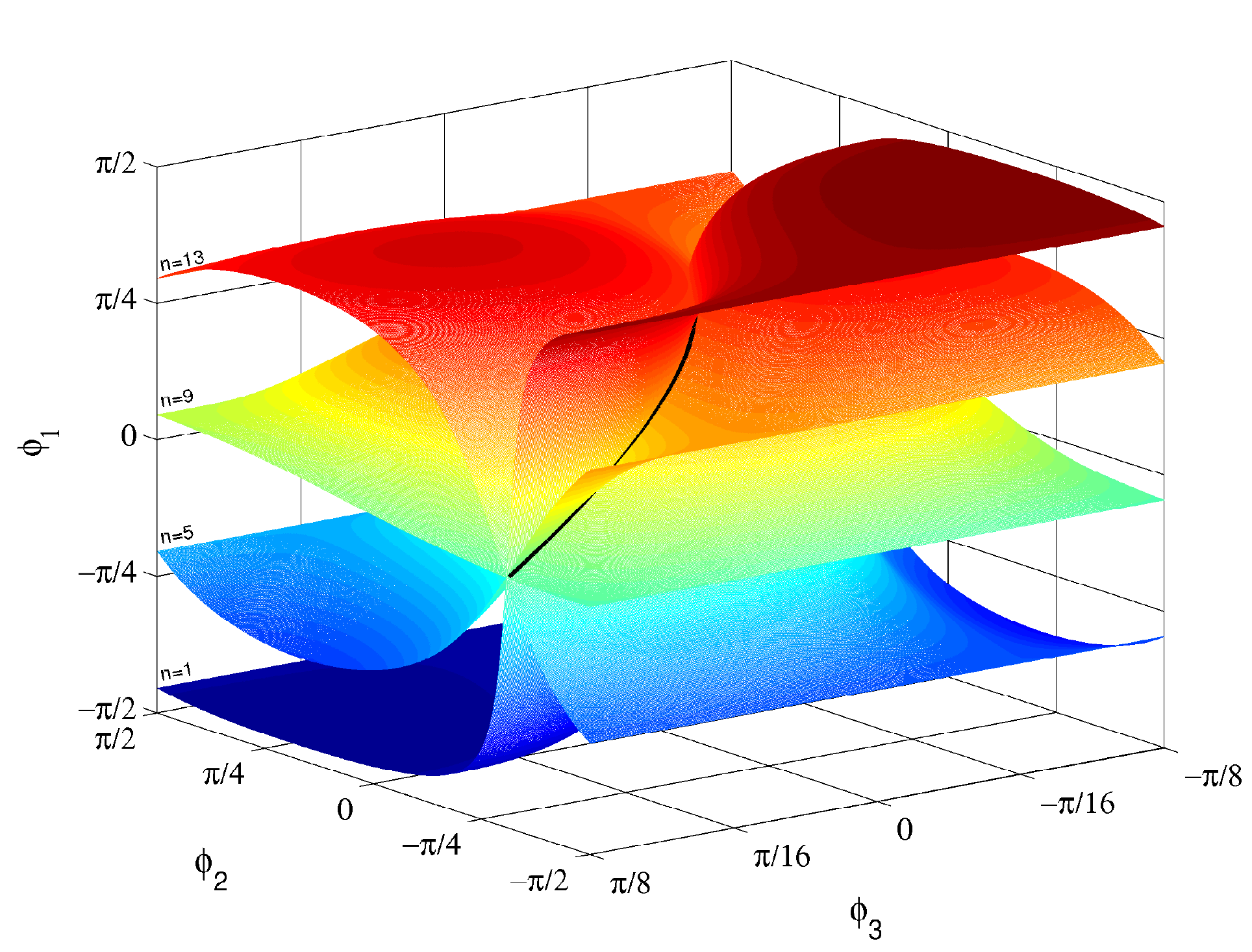}}
\caption{(a) Partition of $ (-\frac{\pi}{2}, \frac{\pi}{2}]^2 \times (-\frac{\pi}{8}, \frac{\pi}{8}] $ using 3 different surfaces $ \mathcal{H}(\tilde{\mathbf{V}}_{1,1:4}) $, $ \mathcal{H}(\tilde{\mathbf{V}}_{2,1:4}) $, and $ \mathcal{H}(\tilde{\mathbf{V}}_{3,1:4}) $.
(b) Intersection of $ \frac{M}{2}=4 $ surfaces that originate from the same row of observation matrix $ \mathbf{V}_{4 \times 2} $.}
\label{figs:1}
\end{figure*}

In~(\ref{eq_010}), we observe that, for any vector $\boldsymbol{\phi}$, the original maximization problem in~(\ref{eq:P}) is decomposed into $N$ symbol-by-symbol maximization rules, according to~\eqref{eq:decision}, and the $n$th maximization argument (i.e., symbol $s_n$) of the sum equals $d(\mathbf{V}_{n,:};\boldsymbol{\phi})=s_n$, that is, it depends only on the $n$th row of $\mathbf{V}$, $n=1,2,\ldots,N$.
As $\boldsymbol{\phi}$ varies, the decision in favor of $ s_n $ is maintained as long as a decision boundary is not crossed.
Due to the structure of $\mathcal{A}_M$, the $\frac{M}{2}$ decision boundaries for the determination of $s_n$ from $\mathbf{V}_{n,:}\mathbf{c}\left(\boldsymbol{\phi}\right)$ are lines that pass through the origin of the complex plane and separate it into disjoint sets and are given by
\begin{equation}
\begin{split}
\mathcal{B}_k^{(n)}\eqdef\Bigl\{&\boldsymbol{\phi}\in\Phi^{2D-2}\times\left(-\frac{\pi}{M},\frac{\pi}{M}\right]:\\
&\mathbf{V}_{n,:} \mathbf{c}(\boldsymbol{\phi}) = A e^{j\pi\frac{2k + 1}{M}}, \;\; A \in \mathbb{R}\Bigr\},\\
k=0,\mbox{}&1,\ldots,\frac{M}{2}-1,\;\;n=1,2,\ldots,N,
\end{split}
\end{equation}
or, equivalently,
\begin{equation}
\begin{split}
&\Im\left\{e^{-j\pi\frac{2k+1}{M}}\mathbf{V}_{n,:}\mathbf{c}(\boldsymbol{\phi})\right\}= 0,\\
&k=0,1,\ldots,\frac{M}{2}-1,\;\;n=1,2,\ldots,N.
\end{split}
\label{eq_011}
\end{equation}
We define
\begin{equation}
\hat{\mathbf{V}}_{\frac{MN}{2}\times D}\eqdef%
\begin{bmatrix}%
e^{-j\pi\frac{1}{M}}\\
e^{-j\pi\frac{3}{M}}\\
\vdots \\
e^{-j\pi\frac{M-1}{M}}
\end{bmatrix}%
\otimes\mathbf{V}.%
\label{eq_012}
\end{equation}
Then, for $ n = 1,2,\dots, N $ and $ k=0,1,\dots,\frac{M}{2}-1 $, we can write (\ref{eq_011}) in matrix form
\begin{equation}
\begin{split}
&\Im\left\{\hat{\mathbf{V}}\mathbf{c}(\boldsymbol{\phi})\right\}={\bf 0}\\
\Leftrightarrow\mbox{}&\Re\left\{\hat{\mathbf{V}}\right\}\tilde{\mathbf{c}}_{1:2:2D-1}(\boldsymbol{\phi})+\Im\left\{\hat{\mathbf{V}}\right\}\tilde{\mathbf{c}}_{2:2:2D}(\boldsymbol{\phi})={\bf 0}\\
\Leftrightarrow\mbox{}&\tilde{\mathbf{V}}\tilde{\mathbf{c}}(\boldsymbol{\phi})={\bf 0}
\end{split}
\label{eq_013}
\end{equation}
where
\begin{equation}
\begin{split}
\tilde{\bf V}=\Bigl[&\Re\left\{\hat{\bf V}_{:,1}\right\}\;\Im\left\{\hat{\bf V}_{:,1}\right\}\;\Re\left\{\hat{\bf V}_{:,2}\right\}\;\Im\left\{\hat{\bf V}_{:,2}\right\}\\
&\ldots\;\Re\left\{\hat{\bf V}_{:,D}\right\}\;\Im\left\{\hat{\bf V}_{:,D}\right\}\Bigr]_{\frac{MN}{2}\times2D}.
\end{split}
\label{eq:Vtilde}
\end{equation}
From the construction of $\tilde{\bf V}$ in~(\ref{eq_012}) and~(\ref{eq:Vtilde}), we observe that the $n$th row of ${\bf V}$ is rotated by each of the $\frac{M}{2}$ exponential terms $e^{-j\pi\frac{2k+1}{M}}$ that represent the decision boundaries $\mathcal{B}_k^{(n)}$, $k=0,1,\ldots,\frac{M}{2}-1$, $n=1,2,\ldots,N$.
Hence, the system of $\frac{MN}{2}$ equations $\tilde{\bf V}\tilde{\bf c}(\boldsymbol{\phi})={\bf 0}$ defines $\frac{MN}{2}$ decision boundaries, such that, as long as a boundary is not crossed by varying $\boldsymbol{\phi}$, the decision $\mathbf{d}(\mathbf{V}; \boldsymbol{\phi})$ remains the same.

\subsection{Hypersurfaces ${\mathcal H}(\tilde{\mathbf{V}}_{i,:})$ and Cardinality of $ \mathbf{\mathcal{S}}(\mathbf{V})$}

According to~(\ref{eq_013}), we can derive $ \frac{MN}{2} $ different decision rules that separate $\Phi^{2D-2}\times\left(-\frac{\pi}{M},\frac{\pi}{M}\right]$ into distinct regions, each of which is associated with a different candidate $M$PSK sequence $\mathbf{s}$.
More specifically, the rows of $\tilde{\mathbf{V}}$ determine $\frac{MN}{2}$ hypersurfaces $\mathcal{H}(\tilde{\mathbf{V}}_{1,:})$, $\mathcal{H}(\tilde{\mathbf{V}}_{2,:})$, $\ldots$, $\mathcal{H}(\tilde{\mathbf{V}}_{\frac{MN}{2},:})$, that is, $(2D-2)$-manifolds in the $(2D-1)$-dimensional space, that partition the $(2D-1)$-dimensional hypercube $\Phi^{2D-2}\times\left(-\frac{\pi}{M},\frac{\pi}{M}\right]$ into $K$ nonoverlapping cells $C_1$, $C_2$, $\ldots$, $C_K$;
the union of all cells is equal to $\Phi^{2D-2}\times\left(-\frac{\pi}{M},\frac{\pi}{M}\right]$ and the intersection of any two distinct cells is empty.
Each cell $C_k$ corresponds to a \textit{distinct} candidate sequence $\mathbf{s}_k\in\mathcal{A}_M^N $ in the sense that $\mathbf{d}(\mathbf{V};\boldsymbol{\phi})=\mathbf{s}_k$ for any $\boldsymbol{\phi}\in C_k$ and $\mathbf{s}_k\neq\mathbf{s}_j$ if $k\neq j$, $k,j\in\{1,2,\ldots,K\}$.

Before we present some further results on the behavior of such hypersurfaces, it is illustrative to present some partitions of $ \Phi^{2D-2} \times (-\frac{\pi}{M}, \frac{\pi}{M}] $ for various values of $D$, $M$, and $N$.
As a first example, we set $D=2$, $N=4$, and $M=8$ and draw an arbitrary $4\times2$ complex matrix $\mathbf{V}$ with $\mathbf{V}_{n,:}\neq\mathbf{0}$, $n=1,2,3,4$.
Since $D=2$ and $M=8$, we are interested only in cells that belong to the region $\boldsymbol{\phi}\in\Phi^2\times\left(-\frac{\pi}{8},\frac{\pi}{8}\right]$.
According to the decision boundary rule in~(\ref{eq_013}), in Fig.~\ref{figs:1}(a), we plot surface $\mathcal{H}(\tilde{\mathbf{V}}_{1,:})$, described by the expression $\phi_1=\tan^{-1}\left(-\frac{\tilde{\mathbf{V}}_{1,2:4}\tilde{\mathbf{c}}(\boldsymbol{\phi}_{2:3})}{\tilde{V}_{1,1}}\right)=\tan^{-1}\left(-\frac{\tilde{V}_{1,2}\sin\phi_2+\tilde{V}_{1,3}\cos\phi_2\sin\phi_3+\tilde{V}_{1,4}\cos\phi_2\cos\phi_3}{\tilde{V}_{1,1}}\right)$ that originates from the first row of $\mathbf{V}$ (depicted as surface $n=1$).
In the same figure, we add two more surfaces, $\mathcal{H}(\tilde{\mathbf{V}}_{2,:})$ and $\mathcal{H}(\tilde{\mathbf{V}}_{3,:})$, that originate from the second and third rows of $\mathbf{V}$ and are denoted as $n=2$ and $n=3$, respectively.
We observe that the surfaces intersect at a single point $\boldsymbol{\phi}(\mathbf{V};\{1,2,3\})$ and the three-dimensional space is partitioned into regions (cells) each of which corresponds to a distinct candidate $M$PSK sequence $\mathbf{s}\in\mathcal{S}(\mathbf{V})$.%
\footnote{For visualization purposes, we do not plot the complete partition.} 

Two basic properties of such intersections are presented in the following proposition.
The proof is given in the Appendix.

\begin{prop}
\label{prop1}
Let $\tilde{\mathbf{V}}_{\frac{MN}{2}\times2D}$ be a real matrix constructed from a $N\times D$ complex matrix $\mathbf{V}$ with $\mathbf{V}_{n,:}\neq\mathbf{0}$, $n=1,2,\ldots,N$.
Then, each subset of $\left\{\mathcal{H}(\tilde{\mathbf{V}}_{1,:}),\mathcal{H}(\tilde{\mathbf{V}}_{2,:}),\ldots,\mathcal{H}(\tilde{\mathbf{V}}_{\frac{MN}{2},:})\right\}$ that consists of $2D-1$ hypersurfaces has
\begin{enumerate}[(a)]
\item[{\it(i)}]
either a single or uncountably many intersections in $\Phi^{2D-2}\times\left(-\frac{\pi}{M},\frac{\pi}{M}\right]$,
\item[{\it(ii)}]
a unique intersection point that constitutes a vertex of a cell if and only if no more than two hypersurfaces originate from the same row of $\mathbf{V}$.
\hfill$\Box$
\end{enumerate} 
\end{prop}

Let $\mathcal{I}\eqdef\left\{i_1,i_2,\ldots,i_{2D-1}\right\}\subset\left\{1,2,\ldots,\frac{MN}{2}\right\}$ denote the subset of $ 2D-1 $ indices that correspond to hypersurfaces $ \mathcal{H}(\tilde{\mathbf{V}}_{i_1,:}), \mathcal{H}(\tilde{\mathbf{V}}_{i_2,:}),\ldots,\mathcal{H}(\tilde{\mathbf{V}}_{i_{2D-1},:})$, respectively.
We identify the following cases.
\begin{enumerate}[(i)]
\item[{\it{(a)}}] Intersections of $ 2D-1 $ hypersurfaces where at most two surfaces originate from the same row of $ \mathbf{V} $.
\item[{\it{(b)}}] Intersections of $ 2D-1 $ hypersurfaces where at least three surfaces originate from the same row of $ \mathbf{V} $.
\end{enumerate}
According to Proposition~\ref{prop1}, Part (ii), combinations in case (b) do not have a unique intersection point but infinitely many intersection points; thus no cell is created and these combinations can be ignored.
A very important observation for our subsequent developments is presented in the following corollary.

\begin{corollary}
All $\frac{M}{2}$ hypersurfaces that originate from the same row of $ \mathbf{V} $ intersect at a common axis.
\hfill$\Box$
\end{corollary}

Extending the previous example with the $4\times2$ matrix ${\bf V}$, we present in Fig.~\ref{figs:1}(b) the intersection of the $\frac{M}{2}=4$ surfaces that originate from the first row of $\mathbf{V}$ and are related with the decision in favor of $s_1$.
In general, such an ensemble of $\frac{M}{2}$ hypersurfaces that originate from the $n$th row of the $N\times D$ matrix ${\bf V}$ partitions the hypercube $\Phi^{2D-2} \times (-\frac{\pi}{M}, \frac{\pi}{M}]$ into $M$ regions, each of which is mapped to a unique element $s_n\in\mathcal{A}_M$.
According to Corollary 1, all $\frac{M}{2}$ hypersurfaces have uncountably many intersection points that form a common axis, that is, a common $(2D-3)$-manifold in the $(2D-1)$-dimensional space.
Thus, hypersurfaces that come from the same row of $\mathbf{V}$ intersect at a common $1$-manifold in the $3$-dimensional space for $D=2$, at a common $3$-manifold in the $5$-dimensional space for $D=3$, etc.%
\footnote{For $D\geq3$, we cannot visualize the resulting partitions and the common intersection $(2D-3)$-manifold.}

On the other hand, combinations in case {\it{(a)}} have a unique intersection point $\boldsymbol{\phi}(\mathbf{V};\mathcal{I})\in\Phi^{2D-2}\times(-\frac{\pi}{M},\frac{\pi}{M}]$
that leads $Q(\mathcal{I})$ cells, say $C_1(\mathbf{V};\mathcal{I}),C_2(\mathbf{V};\mathcal{I}),\ldots,C_{Q(\mathcal{I})}(\mathbf{V};\mathcal{I})$ where $Q(\mathcal{I})\in\left\{\left(\frac{M}{2}-1\right)^0,\left(\frac{M}{2}-1\right)^1,\ldots,\left(\frac{M}{2}-1\right)^{D-1}\right\}$,
and each cell is associated with a distinct candidate $M$PSK sequence $\mathbf{s}_q(\mathbf{V};\mathcal{I})$, in the sense that $\mathbf{s}_q(\mathbf{V};\boldsymbol{\phi})=\mathbf{s}_q(\mathbf{V};\mathcal{I})$ for all $\boldsymbol{\phi}\in C_q(\mathbf{V};\mathcal{I})$ and $\boldsymbol{\phi}(\mathbf{V};\mathcal{I})$ is a single point of $C_q(\mathbf{V};\mathcal{I})$ where $\phi_{2D-1}$ is minimized, $q=1,2,\ldots,Q(\mathcal{I})$.
We underline that there are combinations of hypersurfaces that do not intersect into the region of interest $\Phi^{2D-2}\times(-\frac{\pi}{M},\frac{\pi}{M}]$ but intersect at a single point $\boldsymbol{\phi}(\mathbf{V};\mathcal{I})$ with $\phi_{2D-1}\notin(-\frac{\pi}{M}, \frac{\pi}{M}] $.
As described later, any such case can be ignored since there always exists a combination of hypersurfaces with $\boldsymbol{\phi}(\mathbf{V};\mathcal{I})\in\Phi^{2D-2}\times(-\frac{\pi}{M},\frac{\pi}{M}]$ which ``leads'' cells associated with equivalent candidate sequences.
The number of cells $Q(\mathcal{I})$ ``led'' by an intersection point depends on the number $p$ of participating pairs of hypersurfaces that originate from the same row of matrix $\mathbf{V}$ and is explicitly given by $\left(\frac{M}{2}-1\right)^p$.

\begin{figure*}[!t]
\centerline{\includegraphics[width=3.5in]{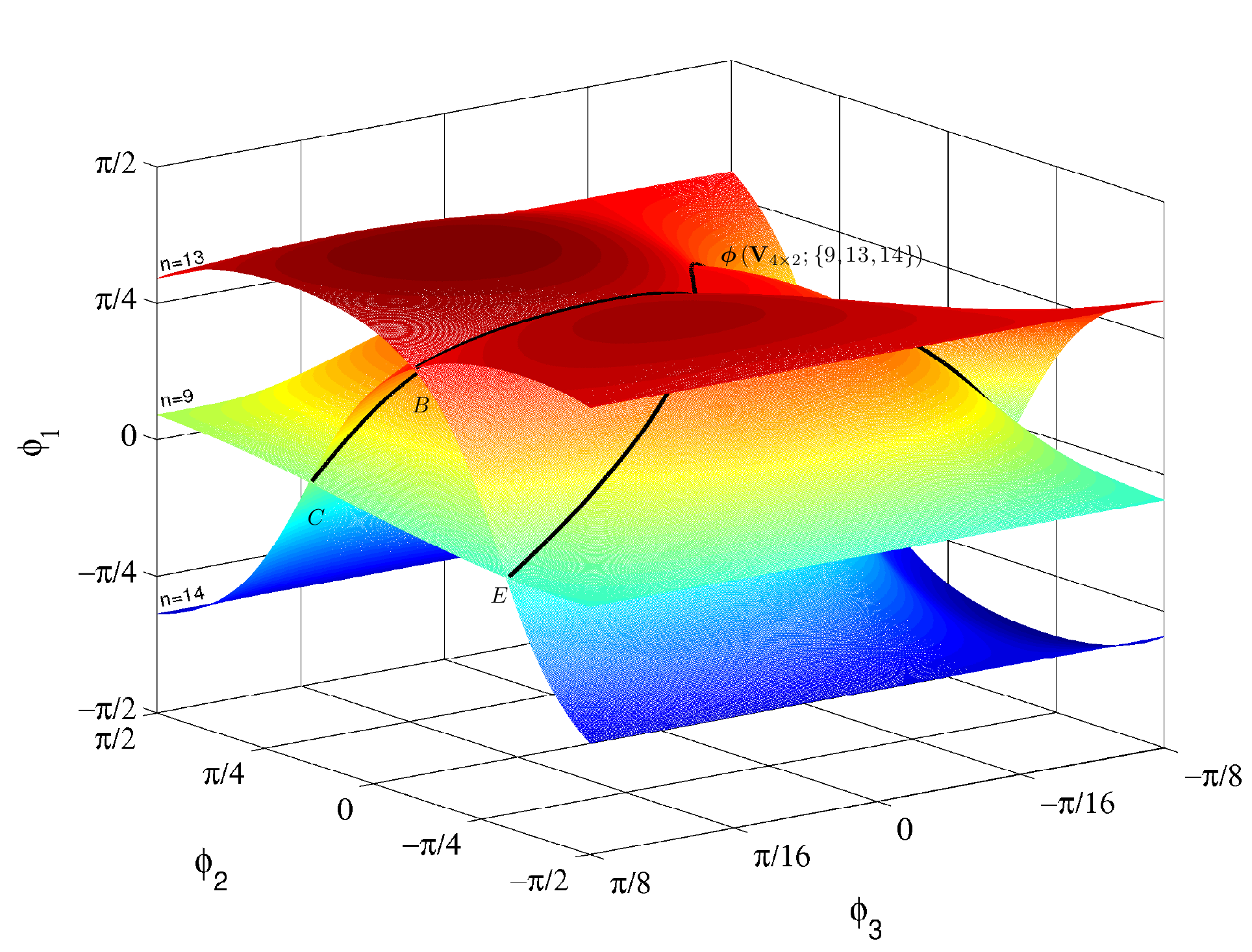}
\hfill
\includegraphics[width=3.5in]{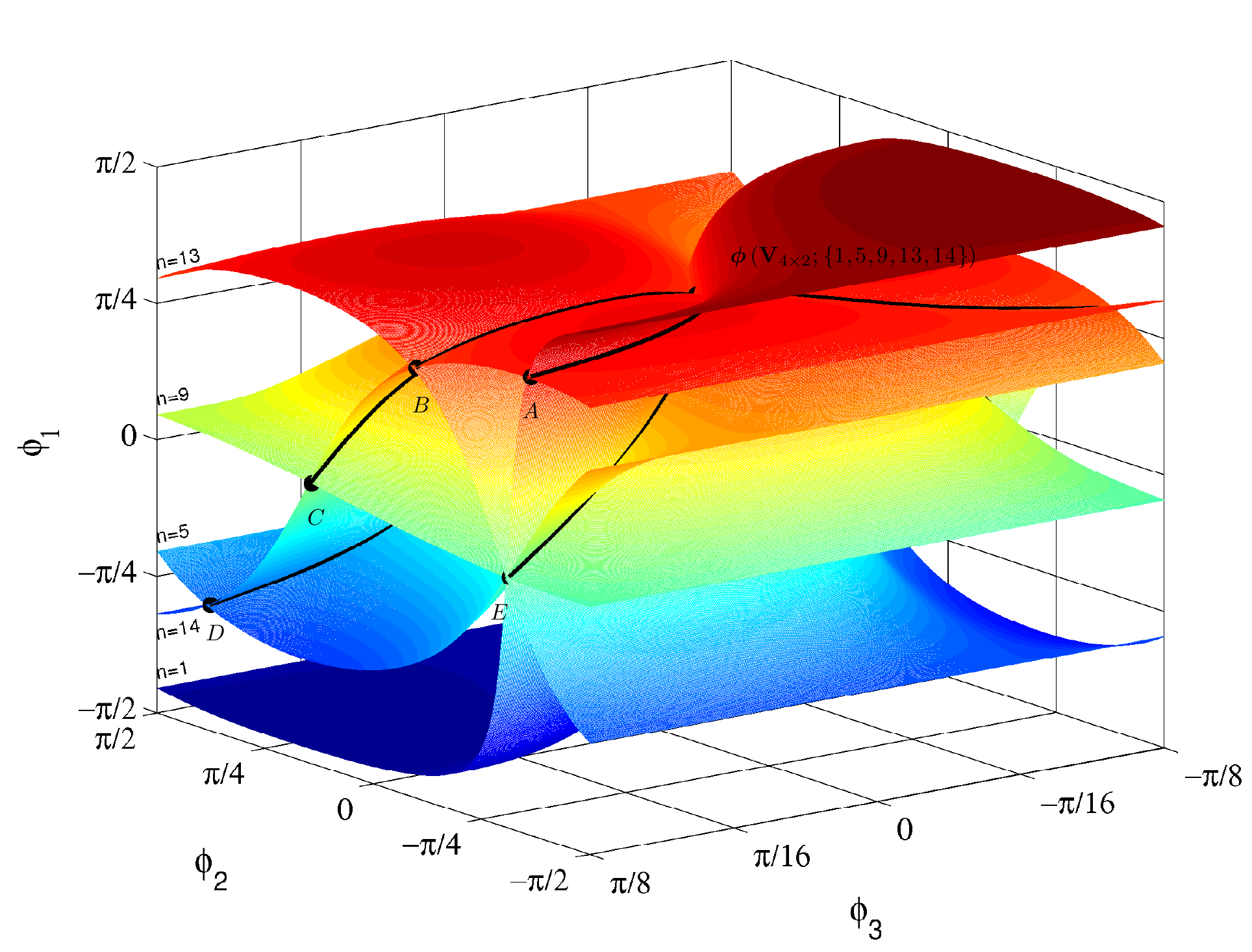}}
\centerline{\includegraphics[width=3.5in]{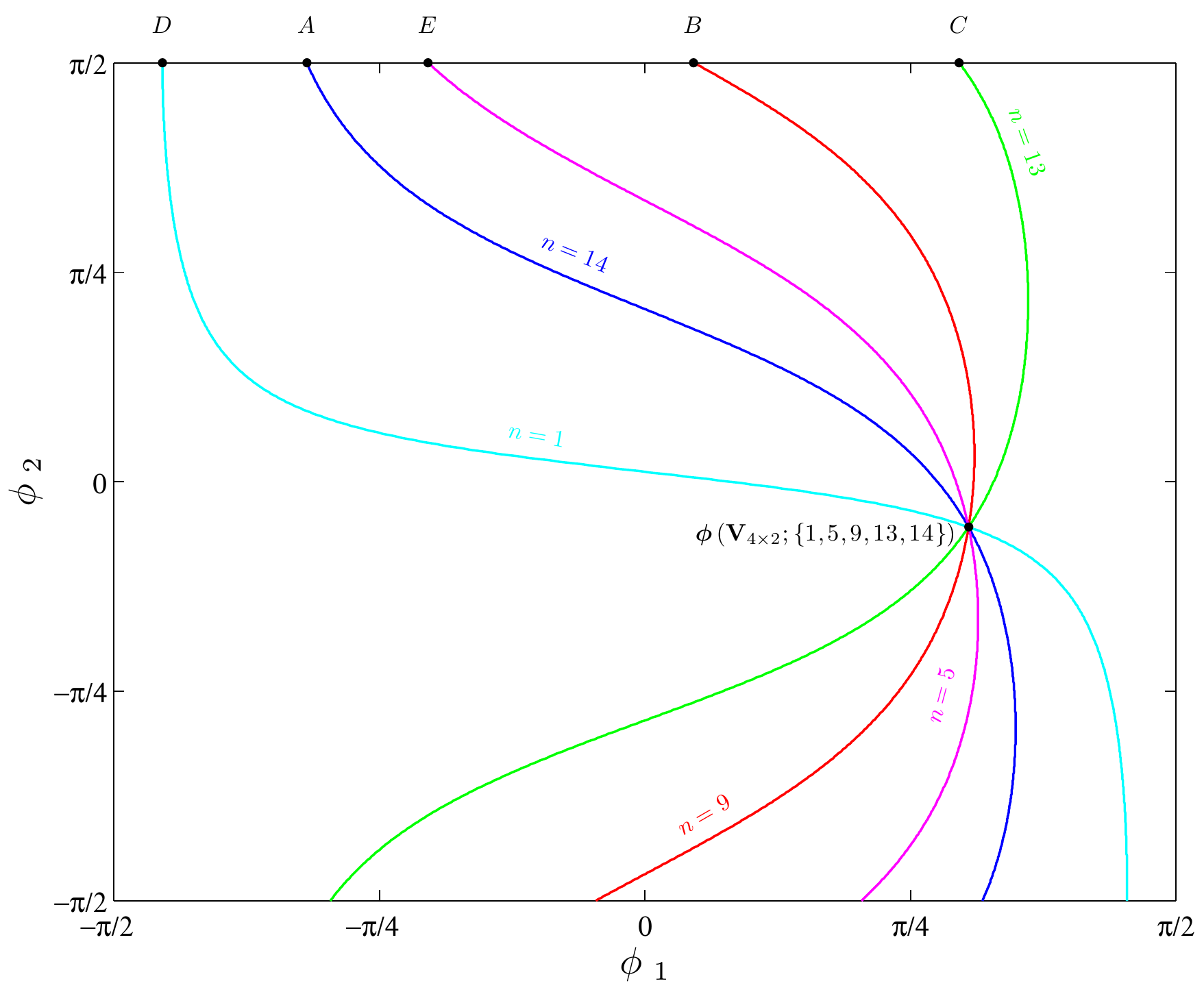}
\hfill
\includegraphics[width=3.5in]{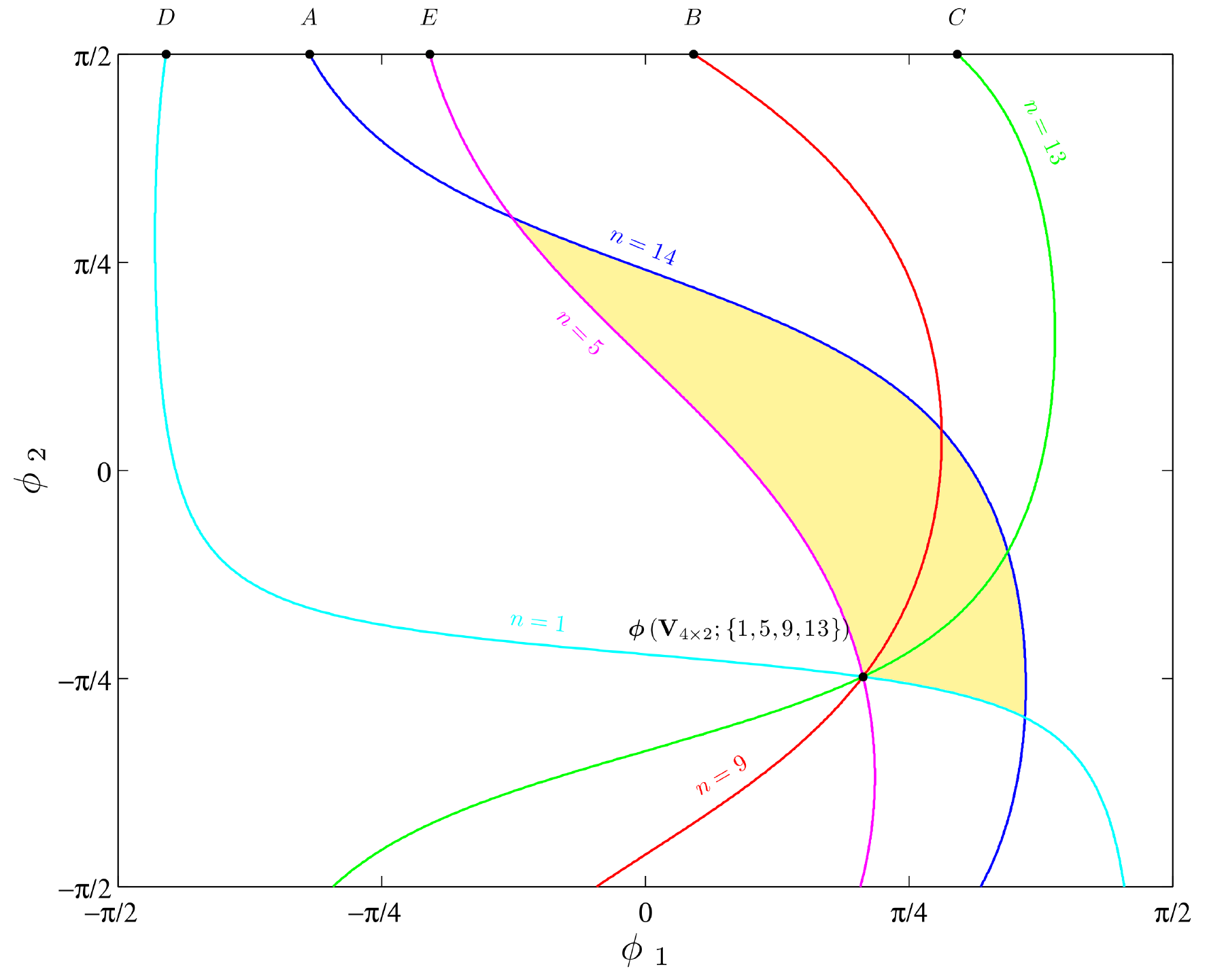}}
\caption{(a) Partition of $ (-\frac{\pi}{2}, \frac{\pi}{2}]^2 \times (-\frac{\pi}{8}, \frac{\pi}{8}] $ using surfaces $ \mathcal{H}(\tilde{\mathbf{V}}_{9,1:4}) $, $ \mathcal{H}(\tilde{\mathbf{V}}_{13,1:4}) $, and $ \mathcal{H}(\tilde{\mathbf{V}}_{14,1:4}) $.
(b) $ \frac{M}{2}=4 $ surfaces, that originate from the first row of $ \mathbf{V}_{4\times2} $, participate in the creation of $\frac{M}{2}-1$ new cells.
(c)-(d) Intersection of surfaces $ \mathcal{H}(\tilde{\mathbf{V}}_{1,1:4}) $, $ \mathcal{H}(\tilde{\mathbf{V}}_{5,1:4}), \mathcal{H}(\tilde{\mathbf{V}}_{9,1:4}) $, $ \mathcal{H}(\tilde{\mathbf{V}}_{13,1:4}) $, and $ \mathcal{H}(\tilde{\mathbf{V}}_{14,1:4}) $ for $ \phi_3 = \arg_{\phi_3} \lbrace \boldsymbol{\phi}(\mathbf{V}_{4 \times 2};\lbrace 9,13,14 \rbrace)\rbrace $ and $ \phi_3 = \arg_{\phi_3} \lbrace \boldsymbol{\phi}(\mathbf{V}_{4 \times 2};\lbrace 9,13,14 \rbrace)\rbrace + d\phi $, respectively.}
\label{figs:2}
\end{figure*}

To better understand the above statements, we illustrate an example with one pair of hypersurfaces originating from the same row of the observation matrix.
For this purpose, we consider the $4\times2$ matrix $\mathbf{V}$ of the previous example where $D=2$, $N=4$, and $M=8$ and present the configuration depicted in Fig.~\ref{figs:2}(a).
Specifically, we present the intersection $\boldsymbol{\phi}(\mathbf{V};\{9,13,14\})$ of two surfaces $\mathcal{H}(\tilde{\mathbf{V}}_{9,:})$ and $\mathcal{H}(\tilde{\mathbf{V}}_{13,:})$ that originate from the first row of $\mathbf{V}$ and one surface $\mathcal{H}(\tilde{\mathbf{V}}_{14,:})$ that originates from the second row of $\mathbf{V}$.
These surfaces are dictated in Fig.~\ref{figs:2} as $n=9$, $n=13$, and $n=14$, respectively.
But, according to Corollary 1, all $\frac{M}{2}$ hypersurfaces that originate from a specific row of $\mathbf{V}$ have a common intersection; thus, in our example, we observe that all $\frac{M}{2}=4$ surfaces $\mathcal{H}(\tilde{\mathbf{V}}_{1,:}),\mathcal{H}(\tilde{\mathbf{V}}_{5,:}),\mathcal{H}(\tilde{\mathbf{V}}_{9,:}),\mathcal{H}(\tilde{\mathbf{V}}_{13,:})$ that originate from the first row of $\mathbf{V}$ pass through the intersection point $\boldsymbol{\phi}(\mathbf{V};\{9,13,14\})=\boldsymbol{\phi}(\mathbf{V};\{1,5,9,13,14\})$; these additional surfaces are dictated as $n=1$ and $n=5$, respectively, in Fig.~\ref{figs:2}(b).
Moving away from the intersection point $\boldsymbol{\phi}(\mathbf{V};\{1,5,9,13,14\})$ by increasing $\phi_3$, the $\frac{M}{2}=4$ surfaces continue to intersect but surface $n=14$ curves away, thus creating $\frac{M}{2}-1$ new cells.%
\footnote{In the sequel, we consider the most computationally demanding case of distinct intersections.}

A better visualization of the above statements is depicted in Figs.~\ref{figs:2}(c)-(d).
For this purpose, we consider the same configuration presented above and present the intersection depicted in Fig.~\ref{figs:2}(a) setting $\phi_3=\arg_{\phi_3}\{\boldsymbol{\phi}(\mathbf{V};\{9,13,14\})\}$.
Thus, in Fig.~\ref{figs:2}(c), we present as functions of $(\phi_1,\phi_2)$ the surfaces that pass through the intersection point $\boldsymbol{\phi}(\mathbf{V};\{9,13,14\})$ where $n=9,$ $n=13$, and $n=14$ are the surfaces under consideration that initially created the intersection point with surfaces $n=9$ and $n=13$ originating from ${\bf V}_{1,:}$ and $n=1$ and $n=5$ denote the remaining surfaces that originate from ${\bf V}_{1,:}$.
According to Corollary 1, we observe that all surfaces pass through the intersection point $\boldsymbol{\phi}(\mathbf{V};\{9,13,14\})$, confirming that $\boldsymbol{\phi}(\mathbf{V};\{9,13,14\})=\boldsymbol{\phi}(\mathbf{V};\{1,5,9,13,14\})$.
In Fig.~\ref{figs:2}(d), we sketch the same surfaces for $\phi_3=\arg_{\phi_3}\{\boldsymbol{\phi}(\mathbf{V};\{9,13,14\})\}+d\phi$ where $d\phi$ is a small arbitrary positive quantity.
We observe that, as $\phi_3$ increases, surface $n=14$ moves away from intersection $\boldsymbol{\phi}(\mathbf{V};\{1,5,9,13,14\})$, thus creating $\frac{M}{2}-1$ cells [see the highlighted cells in Fig.~\ref{figs:2}(d)] that correspond to \textit{distinct} sequences $\mathbf{s}_k\in\mathcal{A}_M^N$.

Since each cell is associated with a distinct candidate $M$PSK sequence, we can collect all these sequences into
\begin{align}
\mathcal{J}(\mathbf{V})\eqdef\hspace{-1.5cm}\bigcup_{\mathcal{I}\subset\left\{1,2,\ldots,\frac{MN}{2}\right\},|\mathcal{I}|=2D-1}\hspace{-1.5cm}\left\{\mathbf{s}_1(\mathbf{V};\mathcal{I}),\mathbf{s}_2(\mathbf{V};\mathcal{I}),\ldots,\mathbf{s}_{Q(\mathcal{I})}(\mathbf{V};\mathcal{I})\right\}\subseteq\mathcal{A}_M^N.
\end{align}
Several properties of the decision function $\mathbf{d}(\mathbf{V};\boldsymbol{\phi})$ are presented in the following proposition.
The proof is provided in the Appendix.

\begin{prop}
\label{prop2}
For any $\boldsymbol{\phi}\in\Phi^{2D-2}\times(-\frac{\pi}{M},\frac{\pi}{M}]$, the following hold true.
\begin{enumerate}[(i)]
\item[{\it{(i)}}]
$\mathbf{d}(\mathbf{V};(\boldsymbol{\phi}_{1:2D-2},-\frac{\pi}{M}))=e^{j\frac{2\pi}{M}}\mathbf{d}(\mathbf{V};(\widehat{\boldsymbol{\phi}}_{1:2D-2},\frac{\pi}{M}))$, for some $\widehat{\boldsymbol{\phi}}_{1:2D-2}\in\Phi^{2D-2}$.
\item[{\it{(ii)}}]
$\mathbf{d}(\mathbf{V};(\boldsymbol{\phi}_{1:2D-3},\frac{\pi}{2},\phi_{2D-1}))=\mathbf{d}(\mathbf{V}_{:,1:D-1};\boldsymbol{\phi}_{1:2D-3})$.
\item[{\it{(iii)}}]
$\mathbf{d}(\mathbf{V};(\boldsymbol{\phi}_{1:2D-3},-\frac{\pi}{2},\phi_{2D-1}))=$\\$-\mathbf{d}(\mathbf{V};(-\boldsymbol{\phi}_{1:2D-3},\frac{\pi}{2},\widehat{\phi}_{2D-1}))$, $\forall\;\widehat{\phi}_{2D-1}\in(-\frac{\pi}{M},\frac{\pi}{M}]$.
\item[{\it{(iv)}}]
$\mathbf{d}(\mathbf{V};(\boldsymbol{\phi}_{1:2D-3},\pm\frac{\pi}{2},\phi_{2D-1}))=$\\$\mathbf{d}(\mathbf{V};(\boldsymbol{\phi}_{1:2D-3},\pm\frac{\pi}{2},\widehat{\phi}_{2D-1}))$, $\forall\;\widehat{\phi}_{2D-1}\in(-\frac{\pi}{M},\frac{\pi}{M}]$.
\hfill$\Box$
\end{enumerate}
\end{prop}

Taking into consideration only cells into the region of interest $ \Phi^{2D-2} \times (-\frac{\pi}{M}, \frac{\pi}{M}] $, we observe that there are
\begin{align}
&|\mathcal{J}(\mathbf{V})|\\
&=\sum_{i=0}^{D-1}\binom{N}{i}\binom{N-i}{2(D-i)-1}\left(\frac{M}{2}\right)^{2(D-i)-2}\left(\frac{M}{2}-1\right)^{i}\nonumber
\end{align}
candidate sequences $\mathbf{s}$ in $\Phi^{2D-2}\times(-\frac{\pi}{M},\frac{\pi}{M}]$ that are associated with cells, each of which minimizes $\phi_{2D-1}$ at a single point that constitutes the intersection of the corresponding $2D-1$ hypersurfaces.
We also note that there exist cells that are not associated with such a vertex and contain uncountably many points of the form $(\phi_1,\ldots,\phi_{2D-2},-\frac{\pi}{M})$.
However, according to Proposition~\ref{prop2}, Part (i), every such a cell can be ignored since there exists another cell that contains points of the form $(\widehat{\phi}_1,\ldots,\widehat{\phi}_{2D-2},\frac{\pi}{M})$, is associated with an equivalent rotated candidate sequence, and is ``led'' by a vertex-intersection that lies in $\Phi^{2D-2}\times(-\frac{\pi}{M},\frac{\pi}{M}]$ (thus, it belongs to $\mathcal{J}(\mathbf{V})$) unless the initial cell contains a point with $\phi_{2D-2}=\pm\frac{\pi}{2}$, as Proposition~\ref{prop2}, Part (iv) mentions.
For example, in Fig.~\ref{figs:1}(a), such cells are identified for $\phi_3=-\frac{\pi}{M}=-\frac{\pi}{8}$.
These candidate $M$PSK sequences are equivalent rotated versions of the sequences determined for $\phi_3=\frac{\pi}{8}$, hence they can be ignored.

In addition, as Proposition~\ref{prop2}, Part (iv) mentions, if $\phi_{2D-2}=\pm\frac{\pi}{2}$ for a particular cell, then this cell ``exists'' for any $\phi_{2D-1}\in(-\frac{\pi}{M},\frac{\pi}{M}]$, implying that we can ignore $\phi_{2D-1}$ (or set it to an arbitrary value $\widehat{\phi}_{2D-1}$), set $\phi_{2D-2}$ to $\pm\frac{\pi}{2}$, and consider cells defined in $\Phi^{2D-3}\times\left\{\pm\frac{\pi}{2}\right\}\times\left\{\widehat{\phi}_{2D-1}\right\}$.
Finally, due to Proposition~\ref{prop2}, Part (iii), the cells that are defined when $\phi_{2D-2}=-\frac{\pi}{2}$ and the cells that are defined when $\phi_{2D-2}=\frac{\pi}{2}$ are associated with opposite sequences.
Therefore, we can ignore the case $ \phi_{2D-2} = -\frac{\pi}{2} $, set $ \phi_{2D-2} = \frac{\pi}{2} $, ignore $ \phi_{2D-1} $, and, according to Proposition~\ref{prop2}, Part (ii), identify the cells that are determined by the \textit{reduced-size} $N\times(D-1)$ matrix $\mathbf{V}_{:,1:D-1}$ in the hypercube $\Phi^{2D-4}\times(-\frac{\pi}{M},\frac{\pi}{M}]$.
For example, in Fig.~\ref{figs:1}(a), we set $\phi_3=\frac{\pi}{8}$ and $\phi_2=\frac{\pi}{2}$ and examine the cells that appear on the leftmost vertical edge of the cube for $\phi_1\in(-\frac{\pi}{8},\frac{\pi}{8}]$.
 
Hence, $\mathcal{S}(\mathbf{V})=\mathcal{J}(\mathbf{V})\cup\mathcal{S}(\mathbf{V}_{:,1:D-1})$ and, by induction,
\begin{equation}
\mathcal{S}(\mathbf{V}_{:,1:d}) = \mathcal{J}(\mathbf{V}_{:,1:d})\cup\mathcal{S}(\mathbf{V}_{:,1:d-1}),\;\;d=2,3,\ldots,D,
\label{eq_019b}
\end{equation}
which implies that
\begin{align}
\mathcal{S}(\mathbf{V})&=\mathcal{J}(\mathbf{V}_{:,1:D})\cup\mathcal{J}(\mathbf{V}_{:,1:D-1})\cup\ldots\cup\mathcal{J}(\mathbf{V}_{:,1:2})\cup\mathcal{J}(\mathbf{V}_{:,1})\nonumber\\
&=\bigcup_{d=0}^{D-1}\mathcal{J}(\mathbf{V}_{:,1:D-d}).
\label{eq_019}
\end{align}
As a result, the cardinality of $\mathcal{S}(\mathbf{V})$ is
\begin{align}
&|\mathcal{S}(\mathbf{V})|=|\mathcal{J}(\mathbf{V}_{:,1:D})|+|\mathcal{J}(\mathbf{V}_{:,1:D-1})|+\ldots+|\mathcal{J}(\mathbf{V}_{:,1:2})|\nonumber\\
&\hspace{1.4cm}+|\mathcal{J}(\mathbf{V}_{:,1})|\nonumber\\
&=\sum_{d = 1}^{D}\sum_{i=0}^{d-1}\binom{N}{i}\binom{N-i}{2(d-i)-1}\left(\frac{M}{2}\right)^{2(d-i)-2}\left(\frac{M}{2}-1\right)^{i}\nonumber\\
&=\mathcal{O}\left(N^{2D-1}\right).\label{eq_prop4_00}
\end{align}

We observe that, if $\mathbf{V}$ is full-rank, i.e., $D=N$, then~(\ref{eq_prop4_00}) returns as many elements as the cardinality of the set $\mathcal{A}_M^{N-1}$, i.e., $\left|\mathcal{S}(\mathbf{V})\right|=\left|\mathcal{A}_M^{N-1}\right|$, as the following proposition states.%
\footnote{In the construction of $\mathcal{S}(\mathbf{V})$, we have, by design, avoided rotated candidate sequences and, thus, the cardinality of the original candidate set drops from $\left|\mathcal{A}_M^N\right|$ to $\left|\mathcal{A}_M^{N-1}\right|$.}
The proof is provided in the Appendix.

\begin{prop}
\label{prop3}
If $D=N$, then $\mathbf{s}_{\text{opt}}$ can be computed through exhaustive search among all elements of $\mathcal{A}_M^{N-1}$, since $\left|\mathcal{S}(\mathbf{V})\right|=\left|\mathcal{A}_M^{N-1}\right|=M^{N-1}$.
\hfill$\Box$
\end{prop}

To summarize the results, we have utilized $2D-1$ auxiliary angles, partitioned the hypercube $\Phi^{2D-2}\times(-\frac{\pi}{M},\frac{\pi}{M}]$ into a finite number of cells that are associated with distinct $M$PSK sequences which constitute the polynomial-size set $\mathcal{S}(\mathbf{V})\subseteq\mathcal{A}_M^N$, and proved that $\mathbf{s}_{\text{opt}}\in\mathcal{S}(\mathbf{V})$.
Therefore, the initial problem in~(\ref{eq:P}) has been converted into numerical maximization of $\left\|\mathbf{V}^{\mathcal{H}}\mathbf{s}\right\|$ among all sequences $\mathbf{s}\in\mathcal{S}(\mathbf{V})$.

\section{Algorithmic Developments}

\subsection{The Proposed Algorithm}

In this section, we present the steps of the proposed algorithm for the construction of $\mathcal{S}(\mathbf{V})$ for arbitrary $N,D\in\mathbb{N}$ with $D\leq N$ and even $M$.
Let $\mathcal{C}_d$ be the set of all combinations of $2d-1$ hypersurfaces that originate from ${\bf V}_{:,1:d}$ and intersect at a single point in $\Phi^{2d-2}\times(-\frac{\pi}{M},\frac{\pi}{M}]$, i.e., $\mathcal{I}=\{i_1,i_2,\ldots,i_{2d-1}\}\in\mathcal{C}_d$ if and only if the intersection of hypersurfaces $\mathcal{H}(\tilde{\mathbf{V}}_{i_1,1:2d}), \mathcal{H}(\tilde{\mathbf{V}}_{i_2,1:2d}), \ldots, \mathcal{H}(\tilde{\mathbf{V}}_{i_{2d-1},1:2d})$ constitutes a vertex of one or more cells in $ \Phi^{2d-2} \times (-\frac{\pi}{M}, \frac{\pi}{M}] $, for $d=1,2,\ldots,D$.
Furthermore, we define $\mathcal{N}_\mathcal{I}\subset\{1,2,\ldots,N\}$ as the set of indices of rows from $ \mathbf{V} $ related with the $ 2d-1 $ hypersurfaces that participate in the intersection point $\boldsymbol{\phi}(\mathbf{V}_{:,1:d};\mathcal{I})$.
From~(\ref{eq_019}), we observe that the initial problem of the determination of $\mathcal{S}(\mathbf{V})$ can be divided into smaller parallel construction problems of $\mathcal{J}(\mathbf{V}_{:,1:d})$, for $d=1,\ldots,D$.
Moreover, the construction of $\mathcal{J}(\mathbf{V}_{:,1:d})$ can be fully parallelized, since the candidate sequence(s) $\mathbf{s}(\mathbf{V}_{:,1:d};\mathcal{I})$ can be computed independently for each $ \mathcal{I}\in\mathcal{C}_d$.
For the following statements, we may assume a certain value for $d\in\{1,2,\ldots,D\}$ and a certain set of indices $\mathcal{I}=\{i_1,i_2,\ldots,i_{2d-1}\}\in\mathcal{C}_d$.
In fact, to simplify notation, we assume that $d=D$ and clarify that similar properties hold also for any $d<D$.

According to the discussion in the previous section, the combination of hypersurfaces $\mathcal{H}(\tilde{\mathbf{V}}_{i_1,:})$, $\mathcal{H}(\tilde{\mathbf{V}}_{i_2,:})$, $\ldots$, $\mathcal{H}(\tilde{\mathbf{V}}_{i_{2D-1},:})$ intersects at a single point $\boldsymbol{\phi}(\mathbf{V};\mathcal{I})$ that ``leads'' the $Q(\mathcal{I})$ cells $C_1(\mathbf{V};\mathcal{I})$, $C_2(\mathbf{V};\mathcal{I})$, $\ldots$, $C_{Q(\mathcal{I})}(\mathbf{V};\mathcal{I})$ associated with $Q(\mathcal{I})$ different candidate $M$PSK sequences $\mathbf{s}_q(\mathbf{V};\mathcal{I})$, $q=1,2,\ldots,Q(\mathcal{I})$.
As already stated, the number of cells $Q(\mathcal{I})$ depends on the number $p$ of pairs of participating hypersurfaces that pass through $\boldsymbol{\phi}(\mathbf{V};\mathcal{I})$ and originate from the same row of $\mathbf{V}$ and equals $\left(\frac{M}{2}-1\right)^{p}$.

To obtain the auxiliary-angle vector $\boldsymbol{\phi}(\mathbf{V};\mathcal{I})$ efficiently, we just need to compute the zero right singular vector of $\tilde{\mathbf{V}}_{\mathcal{I},:}$ and calculate its spherical coordinates.
Specifically, according to the proof of Proposition~\ref{prop1}, Part (i), for a full-rank $(2D-1)\times2D$ real matrix, the system that represents the intersection of $\mathcal{H}(\tilde{\bf V}_{i_1,:})$, $\mathcal{H}({\bf V }_{i_2,:})$, $\ldots$, $\mathcal{H}(\tilde{\bf V}_{i_{2D-1},:})$, i.e.,
\begin{equation}
\tilde{\bf V}_{{\mathcal I},:}\tilde{\mathbf{c}}(\boldsymbol{\phi})={\bf 0},
\end{equation}
has a unique solution $\boldsymbol{\phi}(\mathbf{V};\mathcal{I})\in\Phi^{2D-2}\times(-\frac{\pi}{M},\frac{\pi}{M}]$ which consists of the spherical coordinates of the zero right singular vector of $\tilde{\bf V}_{{\mathcal I},:}$.
Therefore, to obtain $\boldsymbol{\phi}({\bf V};{\mathcal I})$, we just need to compute the zero right singular vector of $\tilde{\bf V}_{{\mathcal I},:}$ and calculate its spherical coordinates.

To identify ${\bf s}_q({\bf V};{\mathcal I})$, $q=1,2,\ldots,Q(\mathcal{I})$, we detect one-by-one its $ N $ elements separately, according to the following rules.%
\footnote{In the following rules, we denote the $n$th element of $\mathbf{s}_q$ as $s_{q,n}$.
If $Q({\mathcal I})=1$, then we denote the $n$th element of the unique candidate ${\bf s}$ as $s_n$.}
\begin{enumerate}
\item[{\it{(i)}}]
For any $n\in\{1,2,\ldots,N\}-{\mathcal N}_{\mathcal I}$, the corresponding element of the candidate sequence ${\bf s}_q({\bf V};{\mathcal I})$ maintains its value at $\boldsymbol{\phi}(\mathbf{V};\mathcal{I})$, hence it is determined by
\begin{equation}
s_{q,n}(\mathbf{V};\mathcal{I})=d(\mathbf{V}_{n,:};\boldsymbol{\phi}(\mathbf{V};\mathcal{I})),\;q=1,2,\ldots,Q(\mathcal{I}).
\label{eq_020}
\end{equation}
\item[{\it{(ii)}}]
For any $n\in\mathcal{N}_\mathcal{I}$ such that there is {\it only one} hypersurface, say $\mathcal{H}(\tilde{\mathbf{V}}_{i_k,:})$, that is related with the $n$th row of $\mathbf{V}$ and participates in the intersection, the corresponding element of $\mathbf{s}_q(\mathbf{V};\mathcal{I})$ cannot be determined at $\boldsymbol{\phi}(\mathbf{V};\mathcal{I})$.
However, it maintains its value at the intersection of the remaining $2D-2$ hypersurfaces $\mathcal{H}(\tilde{\mathbf{V}}_{i_1,:})$, $\mathcal{H}(\tilde{\mathbf{V}}_{i_2,:})$, $\ldots$, $\mathcal{H}(\tilde{\mathbf{V}}_{i_{k-1},:})$, $\mathcal{H}(\tilde{\mathbf{V}}_{i_{k+1},:})$, $\ldots$, $\mathcal{H}(\tilde{\mathbf{V}}_{i_{2D-1},:})$ with the hyperplane $\phi_{2D-1}=\frac{\pi}{M}$.
Note that, by setting $\phi_{2D-1}=\frac{\pi}{M}$,
\begin{align}
\tilde{\bf c}\left(\boldsymbol\phi\right)&=\left[\begin{array}{c}
\sin\phi_1\\
\cos\phi_1\sin\phi_2\\
\cos\phi_1\cos\phi_2\sin\phi_3\\
\vdots\\
\cos\phi_1\ldots\cos\phi_{2D-3}\sin\phi_{2D-2}\\
\cos\phi_1\ldots\cos\phi_{2D-3}\cos\phi_{2D-2}\sin\frac{\pi}{M}\\
\cos\phi_1\ldots\cos\phi_{2D-3}\cos\phi_{2D-2}\cos\frac{\pi}{M}
\end{array}\right]\nonumber\\
&=\left[\begin{array}{cc}
{\bf I}_{2D-2}&{\bf 0}\\
{\bf 0}&\sin\frac{\pi}{M}\\
{\bf 0}&\cos\frac{\pi}{M}
\end{array}\right]\tilde{\bf c}\left(\boldsymbol\phi_{1:2D-2}\right).
\end{align}
Then, for any $2D$-column matrix $\tilde{\bf V}$,
\begin{equation}
\tilde{\bf V}\tilde{\bf c}\left(\boldsymbol\phi\right)=\tilde{\bf V}\left[\begin{array}{cc}
{\bf I}_{2D-2}&{\bf 0}\\
{\bf 0}&\sin\frac{\pi}{M}\\
{\bf 0}&\cos\frac{\pi}{M}
\end{array}\right]\tilde{\bf c}\left(\boldsymbol\phi_{1:2D-2}\right),
\end{equation}
therefore, $s_{q,n}\left({\bf V};{\mathcal I}\right)$ is determined at the point $\boldsymbol\phi$ whose last coordinate $\phi_{2D-1}$ equals $\frac{\pi}{M}$ and first $2D-2$ coordinates $\phi_1,\phi_2,\ldots,\phi_{2D-2}$ are the spherical coordinates of the zero right singular vector of
\begin{equation}
\tilde{\bf V}_{{\mathcal I}-\{i_k\},:}\left[\begin{array}{cc}
{\bf I}_{2D-2}&{\bf 0}\\
{\bf 0}&\sin\frac{\pi}{M}\\
{\bf 0}&\cos\frac{\pi}{M}
\end{array}\right].
\end{equation}
\item[{\it{(iii)}}]
For any $n\in\mathcal{N}_{\mathcal{I}}$ such that there is a pair of hypersurfaces that both originate from the $n$th row of $\mathbf{V}$ and participate in the intersection,
all $\frac{M}{2}$ hypersurfaces ${\mathcal H}\left(\tilde{\bf V}_{n,:}\right)$, ${\mathcal H}\left(\tilde{\bf V}_{n+N,:}\right)$, ${\mathcal H}\left(\tilde{\bf V}_{n+2N,:}\right)$, $\ldots$ , ${\mathcal H}\left(\tilde{\bf V}_{n+\left(\frac{M}{2}-1\right)N,:}\right)$ that originate from the $n$th row of ${\bf V}$ also pass through $\boldsymbol\phi\left({\bf V};{\mathcal I}\right)$.
As a result, $\frac{M}{2}-1$ cells are generated at $\boldsymbol\phi\left({\bf V};{\mathcal I}\right)$ due to those hypersurfaces.
For each such cell, the $n$th element $s_{q,n}$ of its corresponding candidate sequence ${\bf s}_q$ cannot be determined at $\boldsymbol\phi\left({\bf V};{\mathcal I}\right)$ but can be uniquely determined for $\phi_{2d-1}=\frac{\pi}{M}$, as follows.
Consider the intersection of the remaining $2d-3$ hypersurfaces with ${\mathcal H}\left(\tilde{\bf V}_{n,:}\right)$ and hyperplane $\phi_{2d-1}=\frac{\pi}{M}$.
This intersection can be evaluated as in case (ii) above.
At this intersection, we have ambiguity between two specific values of $s_{q,n}\in{\mathcal A}_M$ which maximize the detection metric.
If we repeat with ${\mathcal H}\left(\tilde{\bf V}_{n+N,:}\right)$ instead of ${\mathcal H}\left(\tilde{\bf V}_{n,:}\right)$, then we again have ambiguity between two values, exactly one of which is equal to one of the previous two ambiguity values.
The common value is the actual value of $s_{q,n}$ that corresponds to the cell between ${\mathcal H}\left(\tilde{\bf V}_{n,:}\right)$ and ${\mathcal H}\left(\tilde{\bf V}_{n+N,:}\right)$.
We work similarly with ${\mathcal H}\left(\tilde{\bf V}_{n+N,:}\right)$ and ${\mathcal H}\left(\tilde{\bf V}_{n+2N,:}\right)$, ${\mathcal H}\left(\tilde{\bf V}_{n+2N,:}\right)$ and ${\mathcal H}\left(\tilde{\bf V}_{n+3N,:}\right)$, etc.
This way, we resolve the ambiguity with respect to the $n$th element of the candidate sequences that correspond to the $\frac{M}{2}-1$ cells that are generated due to the $n$th row of ${\bf V}$.
\end{enumerate}

The above statements suggest the following construction of $\mathbf{s}_q(\mathbf{V};\mathcal{I})$, $q=1,2,\dots,Q\left(\mathcal{I}\right)$.
Assuming distinct intersections of hypersurfaces, the $2D-1$ participating hypersurfaces $\mathcal{H}(\tilde{\mathbf{V}}_{i_1,:})$, $\mathcal{H}(\tilde{\mathbf{V}}_{i_2,:})$, $\ldots$, $\mathcal{H}(\tilde{\mathbf{V}}_{i_{2D-1},:})$ pass through the ``leading'' vertex $ \boldsymbol{\phi}(\mathbf{V};\mathcal{I})$ of cell $C_q(\mathbf{V};\mathcal{I})$.
If $n\in\lbrace 1,2, \dots, N \rbrace -\mathcal{N}_{\mathcal{I}} $, i.e., none of the $2D-1$ hypersurfaces originates from the $n$th row of ${\bf V}$, then none of the hypersurfaces that originate from ${\bf V}_{n,:}$ passes through $\boldsymbol{\phi}({\bf V};{\mathcal I}) $. 
As a result, the value of the corresponding $M$PSK element $s_{q,n}(\mathbf{V};\mathcal{I})$ is well determined at the ``leading'' vertex, as (\ref{eq_020}) states.
For example, considering the previous $4\times2$ matrix ${\bf V}$, in Fig.~\ref{figs:1}(a), $s_4({\bf V};\{1,2,3\})$ is well determined at $\boldsymbol{\phi}({\bf V};\{1,2,3\})$ through~(\ref{eq_020}) and maintains its value in the associated cell $C({\bf V};\{1,2,3\})$.

On the other hand, if $n\in\mathcal{N}_{\mathcal{I}}$ such that there is only one hypersurface, say $\mathcal{H}(\tilde{\mathbf{V}}_{i_k,:})$, related to the $n$th row of $\mathbf{V}$, then $\mathcal{H}(\tilde{\mathbf{V}}_{i_k,:})$ passes through $\boldsymbol{\phi}(\mathbf{V};\mathcal{I})$ leading to an ambiguous decision about $s(\mathbf{V}_{n,:};\boldsymbol{\phi}(\mathbf{V};\mathcal{I}))$ between two neighboring $M$PSK elements of ${\mathcal A}_M$, separated by the decision boundary ${\mathcal B}_i^{(n)}$, $i\in\lbrace0,1,\ldots,\frac{M}{2}-1\rbrace$, related to $\mathcal{H}(\tilde{\mathbf{V}}_{i_k,:})$.
For example, in Fig.~\ref{figs:1}(a), the hypersurfaces $\mathcal{H}(\tilde{\mathbf{V}}_{1,:})$, $\mathcal{H}(\tilde{\mathbf{V}}_{2,:})$, and $\mathcal{H}(\tilde{\mathbf{V}}_{3,:})$ originate from the first, second, and third, respectively, row of the $4\times2$ matrix $\mathbf{V}$ and pass through $\boldsymbol{\phi}(\mathbf{V};\{1,2,3\})$ leading to ambiguous decisions for $d(\mathbf{V}_{1,:};\boldsymbol{\phi}(\mathbf{V};\{1,2,3\}))$, $d(\mathbf{V}_{2,:};\boldsymbol{\phi}(\mathbf{V};\{1,2,3\}))$, and $d(\mathbf{V}_{3,:};\boldsymbol{\phi}(\mathbf{V};\{1,2,3\}))$, respectively.
In such a case, ambiguity is resolved if we exclude $\mathcal{H}(\tilde{\mathbf{V}}_{i_k,:})$ and consider the intersection of the remaining $2D-2$ hypersurfaces at $\phi_{2D-1}=\frac{\pi}{M}$ where the value of $s_{q,n}$ is well determined and equals the value of $s_{q,n}$ at any point of the cell of interest $C_q(\mathbf{V};\mathcal{I})$.
For example, in Fig.~\ref{figs:1}(a), the ambiguity with respect to $s_1(\mathbf{V};\{1,2,3\})$, $s_2(\mathbf{V};\{1,2,3\})$, and $s_3(\mathbf{V};\{1,2,3\})$ at intersection $\boldsymbol{\phi}(\mathbf{V};\{1,2,3\})$ is resolved at $C=\boldsymbol{\phi}(\mathbf{V};\{2,3\})$, $A=\boldsymbol{\phi}(\mathbf{V};\{1,3\})$, and $B=\boldsymbol{\phi}(\mathbf{V};\{1,2\})$, respectively.

Finally, if $n\in\mathcal{N}_{\mathcal I}$ such that there is a pair of hypersurfaces, say $\mathcal{H}(\tilde{\mathbf{V}}_{i_k,:})$, $\mathcal{H}(\tilde{\mathbf{V}}_{i_m,:})$, originating from the $n$th row of ${\bf V}$, then, according to Corollary 1, all hypersurfaces that originate from ${\bf V}_{n,:}$ pass through $\boldsymbol{\phi}({\bf V};{\mathcal I})$.
Thus, point $\boldsymbol{\phi}(\mathbf{V};{\mathcal I})$ belongs to the common intersection of the $\frac{M}{2}$ hypersurfaces that originate from ${\bf V}_{n,:}$ and, therefore, we have ambiguity for $s_{q,n}$ among all elements of the $M$PSK alphabet (there is no preference between two specific $M$PSK symbols as in case {\it (ii)}).
The ambiguity is resolved if we exclude hypersurfaces $\mathcal{H}(\tilde{\mathbf{V}}_{i_k,:})$, $\mathcal{H}(\tilde{\mathbf{V}}_{i_m,:})$ and compute the intersection point of the remaining $2D-3$ hypersurfaces with each one of the hypersurfaces originating from ${\bf V}_{n,:}$ that ``construct'' the cells $C_q(\mathbf{V};\mathcal{I})$, $q=1,2,\ldots,Q\left({\mathcal I}\right)$ at $\phi_{2D-1}=\frac{\pi}{M}$.
Since each cell $C_q({\bf V};{\mathcal I})$ is ``constructed'' by hypersurfaces that correspond to decision boundaries ${\mathcal B}_k^{(n)}$, $k=0,1,\ldots,\frac{M}{2}-1$, and are obtained be consecutive rotations of ${\bf V}_{n,:}$ according to~(\ref{eq_011}), these intersection points lead to ambiguous decision sets about $s_{q,n}$ between neighboring elements of the $M$PSK alphabet.
The intersection of these sets determines the value of the corresponding $M$PSK element $s_{q,n}({\bf V};{\mathcal I})$ for each cell. 

For example, in Fig.~\ref{figs:2}(a), surfaces $\mathcal{H}(\tilde{\bf V}_{9,:})$ and $\mathcal{H}(\tilde{\bf V}_{13,:})$ ($n=9$ and $n=13$, respectively) originate from the first row of ${\bf V} $ while $\mathcal{H}(\tilde{\bf V}_{14,:})$ ($n=14$) comes from the second row of ${\bf V}$.
Adding the rest hypersurfaces that originate from the first row of ${\bf V}$, we obtain Fig.~\ref{figs:2}(b) where we observe that $\boldsymbol{\phi}(\mathbf{V};\{1,5,9,13,14\})$ ``leads'' $\frac{M}{2}-1$ cells described by points $\{A,B,E,\boldsymbol{\phi}(\mathbf{V};\{1,5,9,13,14\})\}$, $ \{B,C,E,\boldsymbol{\phi}(\mathbf{V};\{1,5,9,13,14\})\}$, and $\{C,D,E,\boldsymbol{\phi}(\mathbf{V};\{1,5,9,13,14\})\}$.
Each one of the aforementioned cells is related with a different candidate $M$PSK sequence $ \mathbf{s}_q(\mathbf{V};\mathcal{I})$.
Taking as example the cell that contains point $\{A,B,E,\boldsymbol{\phi}(\mathbf{V};\{1,5,9,13,14\})\}$, the ambiguity of $s_{1,1}(\mathbf{V};\{1,5,9,13,14\})$ in this cell is resolved by computing the ambiguity decision sets at points $A$ and $B$ with respect to $s_1$ and finding the common $M$PSK element of these sets.
The same procedure is repeated for the other two cells.

A MATLAB code that implements the proposed algorithm following the construction described above is available at http://www.telecom.tuc.gr/$\sim$karystinos.

\subsection{Properties of the Proposed Algorithm}

The proposed algorithm visits independently $|\mathcal{S}(\mathbf{V})|=\mathcal{O}\left((\frac{MN}{2})^{2D-1}\right)$ intersections and computes the candidate $M$PSK sequence(s) associated with each intersection. 
For each ${\mathcal I}_{2d-1}\in{\mathcal C}_d$, the cost of the algorithm is $\mathcal{O}\left(\frac{MN}{2}\right)$, $d=1,2,\ldots,D$.
Therefore, the overall complexity of the algorithm for the computation of $\mathcal{S}(\mathbf{V})$ becomes $\mathcal{O}\left((\frac{MN}{2})^{2D-1}\right)\mathcal{O}\left(\frac{MN}{2}\right)=\mathcal{O}\left((\frac{MN}{2})^{2D}\right)$.
That is, the proposed algorithm has lower complexity than the method in~\cite{CG:00} (which treats only the cases $M=2$ and $M=4$) and the same order of complexity with the algorithms in~\cite{Iran} and~\cite{MLSD:00}.
Moreover, it is superior to the algorithms in~\cite{Iran} and~\cite{MLSD:00} in terms of exact computational cost, parallelizability, and memory efficiency, as explained below.

Regarding computational cost comparisons, the work in~\cite{Iran} does not mention the exact size of the constructed candidate sequence set, but only proves that it is constructed with complexity $\mathcal{O}\left(N^{2D}\right)$.
As explained in Subsection~\ref{subsec:contribution}, the algorithm in~\cite{Iran} examines at least $\binom{MN}{2D-1}$ ``potential'' candidate sequences, implying that $\binom{MN}{2D-1}$ is a lower bound on the computational cost that~\cite{Iran} requires to build $\mathcal{S}(\mathbf{V})$.
This lower bound is clearly higher than the computational cost of the proposed approach, as also indicated in Figs.~\ref{fig:M4} and~\ref{fig:M8}.
We also recall that the work in~\cite{MLSD:00} identifies neither the exact size of $\mathcal{S}(\mathbf{V})$ (or any bounds on it) nor the computational cost (or any bounds on it) to build $\mathcal{S}(\mathbf{V})$.

Regarding parallelizability and memory efficiency, we observe that the computation of the candidate sequences of ${\mathcal S}({\bf V})$ by the proposed algorithm is performed independently from cell to cell, which implies that there is no need to store the data that have been used for each candidate and we only have to store the ``best'' sequence that has been met.
The memory utilization of the proposed method is, therefore, minimized, in contrast to the algorithm in~\cite{MLSD:00} which has large memory requirement because it utilizes the incremental algorithm for cell enumeration in arrangements~\cite{Inc:00},~\cite{Edelsbrunner-book} which is memory inefficient, since it needs to store all the extreme points, all faces, and their incidences.
The fact that the $\left|{\mathcal S}\left({\bf V}\right)\right|$ cells are examined independently of each other (hence, the corresponding candidate sequences are computed independently of each other) implies that the proposed algorithm is also fully parallelizable, in sharp contrast to the algorithm in~\cite{MLSD:00}.

Finally, we mention that, due to~(\ref{eq_019}), the proposed method is rank-scalable.
If the initial problem is of a high rank that makes the optimization intractable, then matrix ${\bf V}$ in~(\ref{eq:P}) can be approximated by keeping its $d$ strongest principal components.
In such a case, the optimization begins with rank $d=1$ and successive principal components are introduced to increase $d$ and, hence, expand ${\mathcal S}({\bf V})$, until a satisfactory reduced-rank approximation is reached.
Interestingly, according to~(\ref{eq_019b}), ${\mathcal S}({\bf V}_{:,1:d})\subset{\mathcal S}({\bf V}_{:,1:d+1})$, for any $d=1,2,\ldots$.
That is, as we increase the rank of the approximation of ${\bf V}$, the new candidates that are generated are added on the previous ones and optimality with respect to ${\mathcal S}({\bf V})$ is maintained.

\section{Simulation Studies}

\subsection{ML Noncoherent $M$PSK Sequence Detection}

We consider the ML noncoherent sequence receiver for $M$PSK signals in a single-input multiple-output (SIMO) system with $D$ receive antennas and unknown channel state information at both transmitter and receiver ends.
Let ${\bf s}\in\mathcal{A}_M^N$ be the transmitted data vector.
The $N\times D$ observation matrix ${\bf Y} \eqdef [{\bf y}_1 \: {\bf y}_2 \: \ldots \: {\bf y}_D]$ contains as columns the $N$-long data streams received by the $D$ corresponding antennas, where
\begin{equation}
{\bf y}_d=h_d{\bf s}+{\bf n}_d
\end{equation}
is the length-$N$ signal vector received by the $d$th antenna, $h_d$ is the corresponding channel coefficient, and ${\bf n}_d\sim{\mathcal{CN}}\left({\bf 0}_{N\times1},\sigma^2{\bf I}_{N\times N}\right)$ accounts for additive white complex Gaussian channel noise, $d=1,2,\ldots,D$.
The optimal decision is given by
\begin{equation}
{\bf s}_\text{opt}\eqdef\argmax_{{\bf s}\in\mathcal{A}^N_M}f({\bf Y}|{\bf s})=\argmax_{{\bf s}\in\mathcal{A}^N_M}f({\bf y}_1,{\bf y}_2,\ldots,{\bf y}_D|{\bf s})
\end{equation}
where $f(\cdot|\cdot)$ represents the pertinent matrix/vector probability density function of the channel output conditioned on a symbol sequence.
Assuming independent and identical Rayleigh distribution for the $D$ flat-fading channels, the ML sequence detector (MLSD) decides in favor of~\cite{MLSD:00}-\cite{STBC}
\begin{equation}
{\bf s}_\text{opt}=\argmax_{{\bf s}\in\mathcal{A}^N_M}\left\|{\bf Y}^\mathcal{H}{\bf s}\right\|^2=\argmax_{{\bf s}\in\mathcal{A}^N_M}\left\{{\bf s}^\mathcal{H}{\bf Y}{\bf Y}^\mathcal{H}{\bf s}\right\}.
\end{equation}
Therefore, ML noncoherent detection in SIMO systems results in the fixed-rank quadratic-form maximization problem in~(\ref{eq:P}) that can be efficiently solved in polynomial time by the proposed algorithm.

\begin{figure}[t]%
\centerline{\includegraphics[width=\columnwidth]{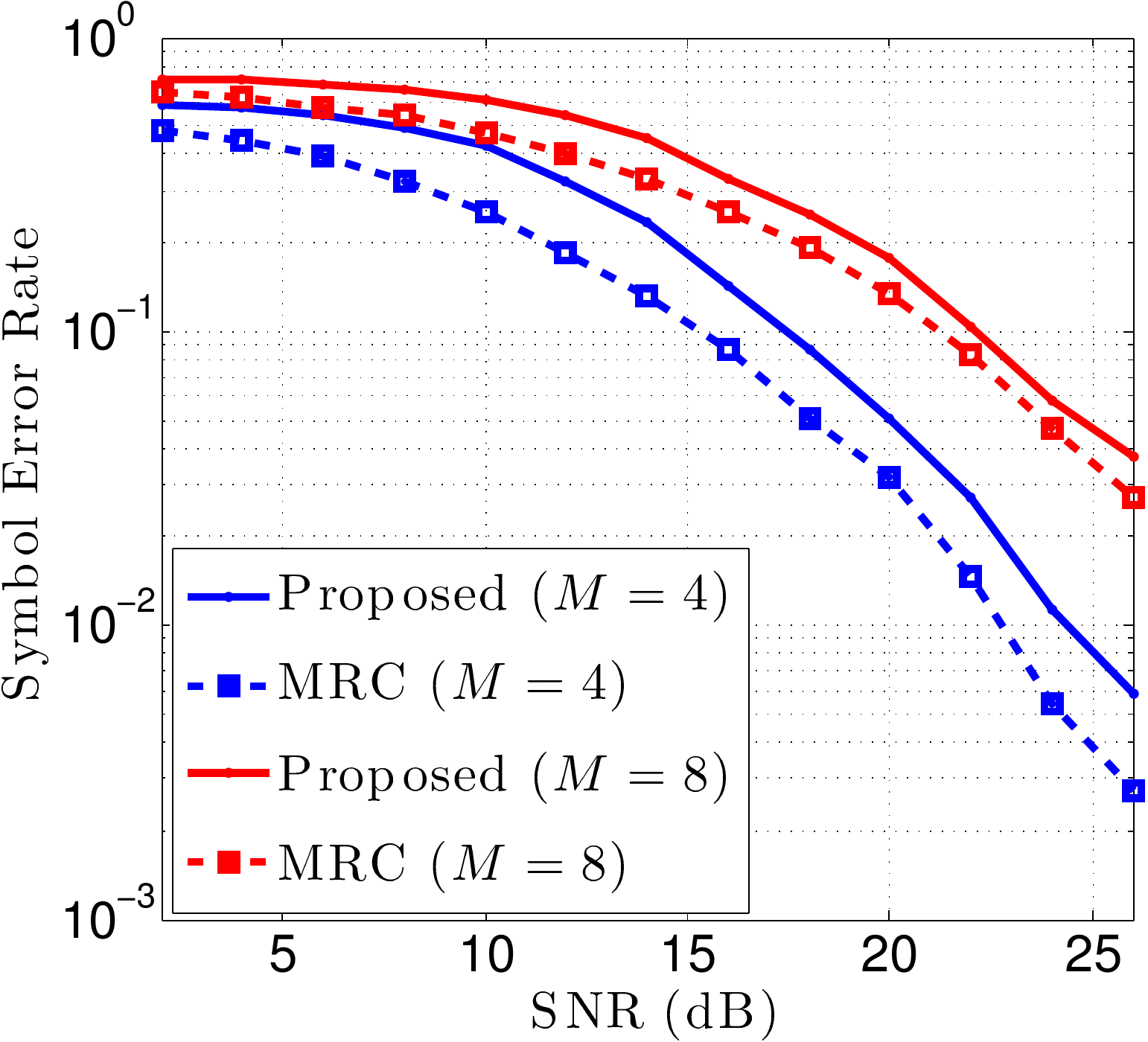}}
\caption{$4$PSK or $8$PSK SER versus SNR of MLSD with sequence length $N=16$ and MRC.}%
\label{fig:SIMO_N16}%
\end{figure}%

\begin{figure*}[t]%
\begin{minipage}{.30\textwidth}%
\centerline{\includegraphics[width=\textwidth]{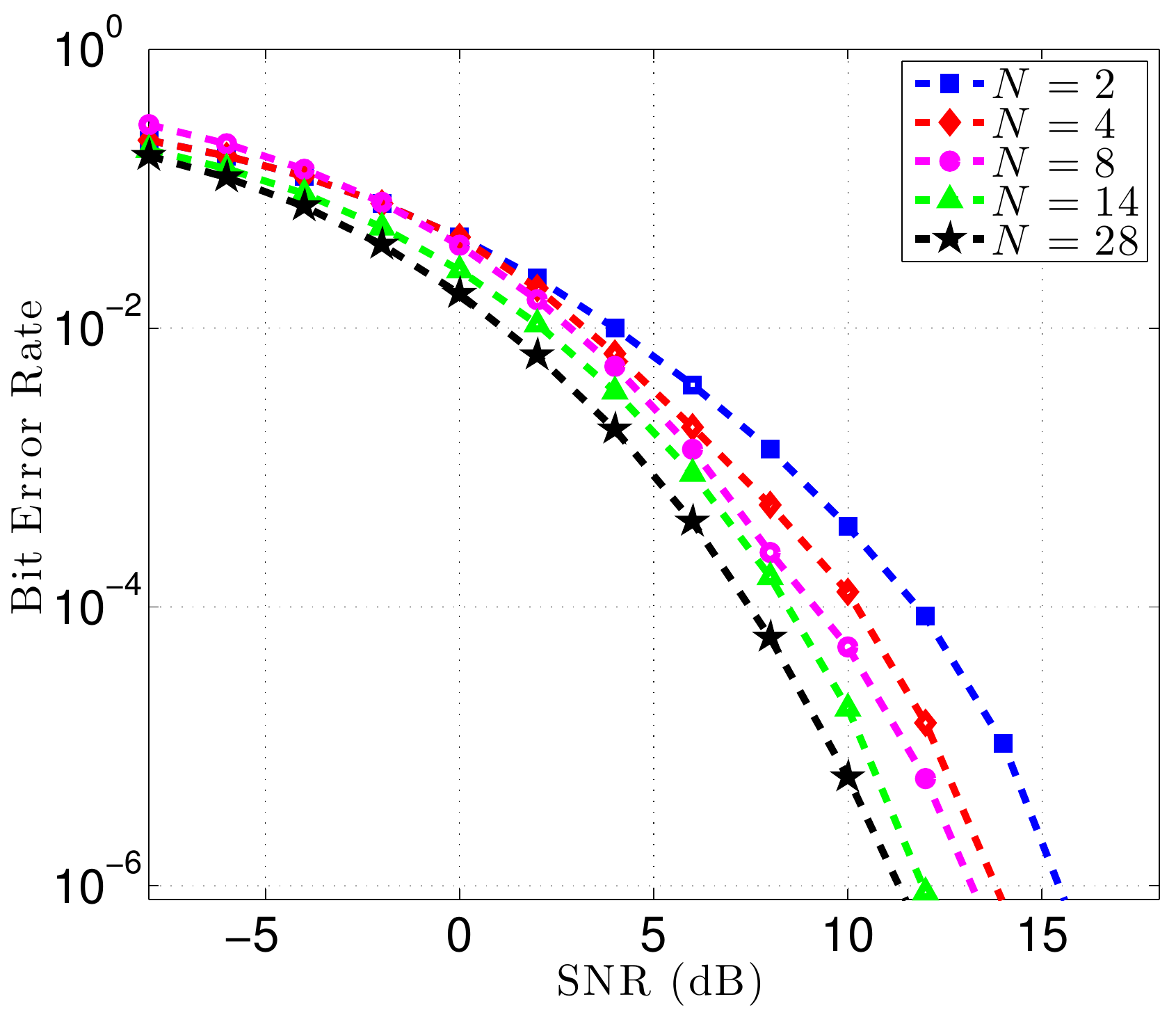}}
\caption{BER versus SNR for $4$PSK beamforming.}%
\label{fig:MIMO_M4}%
\end{minipage}%
\hfill%
\begin{minipage}{.30\textwidth}%
\centerline{\includegraphics[width=\textwidth]{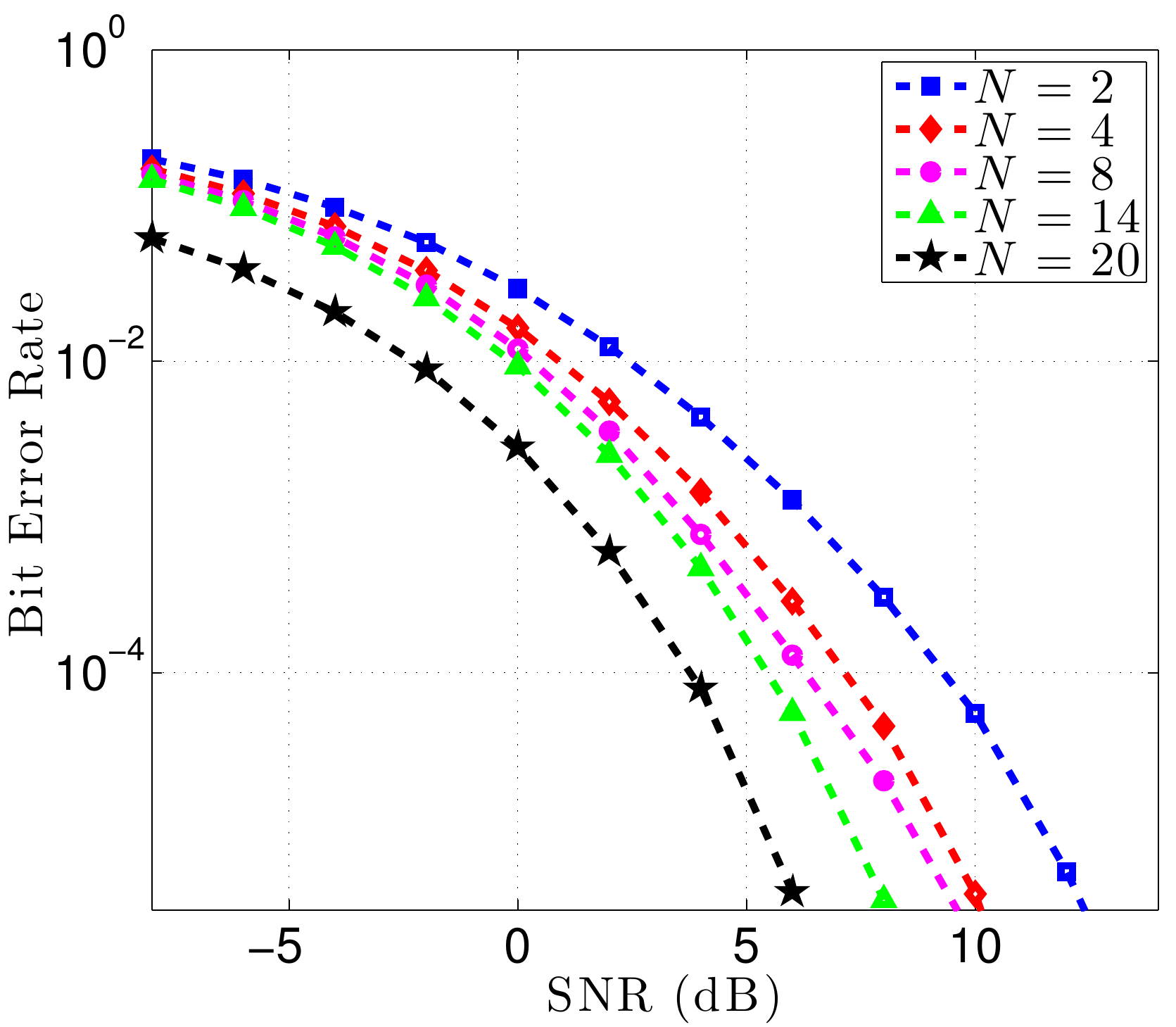}}
\caption{BER versus SNR for $8$PSK beamforming.}%
\label{fig:MIMO_M8}%
\end{minipage}%
\hfill%
\begin{minipage}{.30\textwidth}%
\centerline{\includegraphics[width=\textwidth]{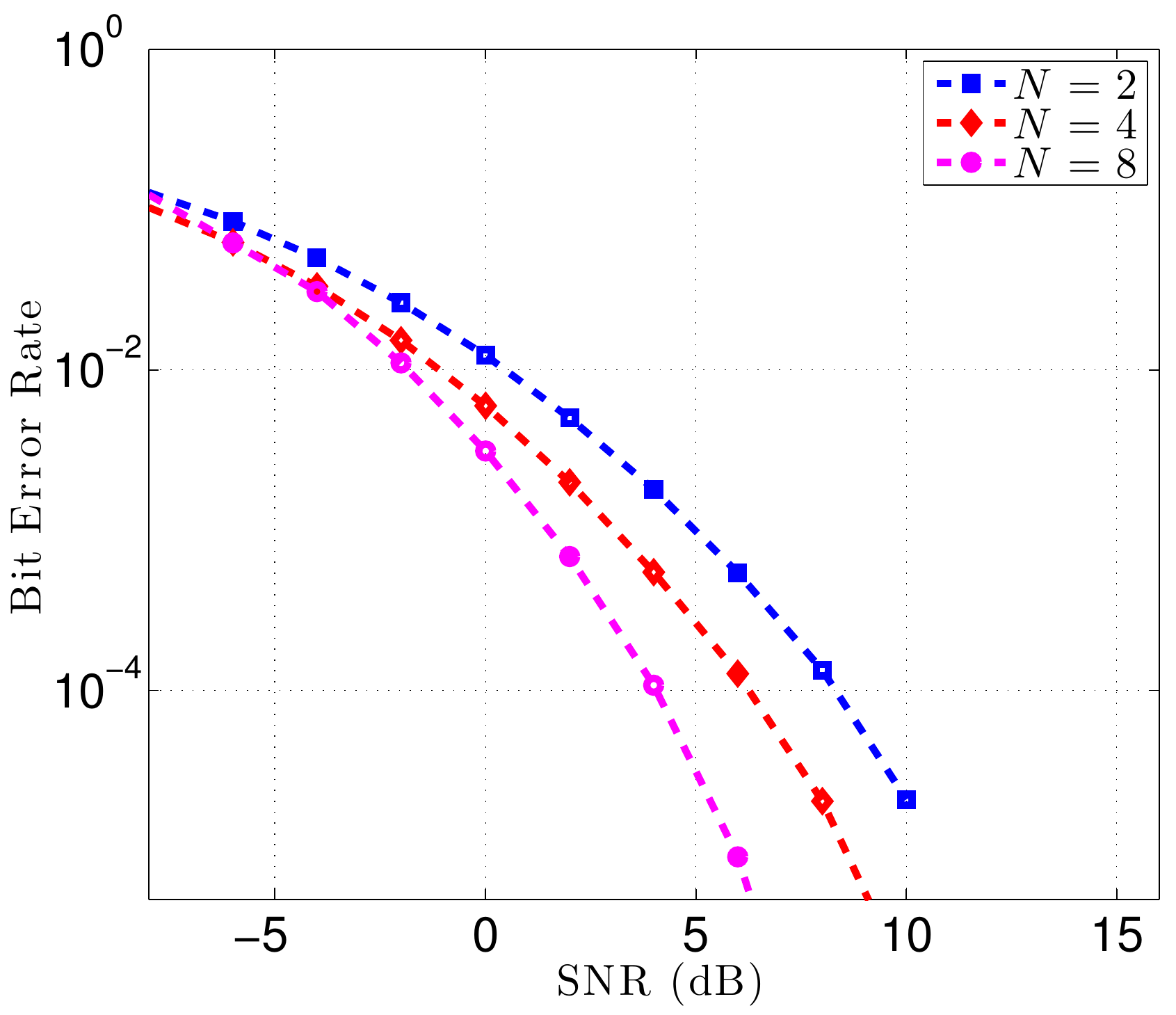}}%
\caption{BER versus SNR for $16$PSK beamforming.}%
\label{fig:MIMO_M16}%
\end{minipage}%
\end{figure*}%

To provide an illustration for the optimal MLSD in SIMO systems, we consider a $1\times2$ SIMO system with $4$PSK ($M=4$) and $8$PSK ($M=8$) transmissions and unknown channel state information at the receiver.
To resolve the phase ambiguity induced by the channel, we utilize differential encoding and perform ML noncoherent sequence detection implemented by the proposed algorithm with polynomial complexity of order ${\mathcal O}\left(N^4\right)$.
The results that we present are averages over $1,000$ randomly generated channel realizations.
In Fig.~\ref{fig:SIMO_N16}, we plot the symbol error rate (SER) of the MLSD receiver for a sequence length $N=16$ as a function of the signal-to-noise ratio (SNR).
As a reference, we include the SER of the maximal ratio combining (MRC) receiver which assumes known channel state information at the receiver.
We emphasize that the MLSD receiver cannot be implemented in reasonably small time through exponential-complexity exhaustive search while the proposed algorithm offers ML performance with polynomial computational complexity.
For example, for the case $M=8$ and $N=16$ in Fig.~\ref{fig:SIMO_N16}, the exhaustive-search MLSD requires a search among $M^{N-1}\simeq3.5\times10^{13}$ vectors of length $16$ while the proposed implementation of MLSD performs a search among $N+\binom{N}{3}\left(\frac{M}{2}\right)^2+N(N-1)\left(\frac{M}{2}-1\right)\simeq10^4$ vectors of length $16$.

\subsection{Limited-feedback Constant-envelope Transmit Beamforming in MIMO Systems}

We consider transmit beamforming on a flat-fading $N\times D$ MIMO communication channel with maximum-SNR filtering at the receiver end.
The number of transmit and receive antennas is $N$ and $D$, respectively.
The $D\times N$ channel matrix is denoted by ${\bf H}$ and assumed to remain stable over some transmission period, such that it is estimated by the receiver.
The received vector is
\begin{equation}
{\bf y}={\bf H}{\bf w}x+{\bf n}
\end{equation}
where ${\bf w}$ is the $N\times1$ beamforming vector, $x$ is the transmitted symbol, and ${\bf n}$ represents additive zero-mean disturbance with covariance matrix ${\bf R}$.
The receiver utilizes the maximum-SNR filter ${\bf R}^{-1}{\bf H}{\bf w}$ to process ${\bf y}$ and the filter-output SNR is
\begin{equation}
E\left\{|x|^2\right\}{\bf w}^\mathcal{H}{\bf H}^\mathcal{H}{\bf R}^{-1}{\bf H}{\bf w}.
\label{eq:SNR}
\end{equation}
The objective is to design ${\bf w}$ to maximize the above expression.

If the receiver has only {\it limited-feedback} capabilities, then the beamforming vector is selected from a predefined $M$PSK codebook (which is due to the fact that usually a per-antenna-element power constraint is enforced at the transmitter, resulting in constant-envelope beamforming).
In particular, the receiver computes the beamforming vector $\mathbf{w}_\text{opt}$ that maximizes the filter-output SNR in~(\ref{eq:SNR}), according to~\cite{Heath1998}-\cite{Santipach2011}
\begin{equation}
\mathbf{w}_{\text{opt}} = \argmax_{{\bf w} \in \mathcal{A}_M^N} \left\{{\bf w}^\mathcal{H}{\bf H}^\mathcal{H}{\bf R}^{-1}{\bf H}{\bf w}\right\}.
\label{eq:wSNR}
\end{equation}
In~(\ref{eq:wSNR}), the $N\times N$ matrix ${\bf H}^\mathcal{H}{\bf R}^{-1}{\bf H}$ has rank $D$.
Therefore, maximum-SNR limited-feedback transmit beamforming over the $M$PSK alphabet results in the quadratic-form maximization problem in~(\ref{eq:P}).
When the number of transmit antennas $N$ is large (as in massive MIMO systems) and the number of receive antennas $D$ is small (if, for example, the receiver is a mobile terminal), the above optimization problem can be efficiently solved in polynomial time by the proposed algorithm.

As an illustration, we consider three $M$PSK beamforming codebooks, for $M = 4$, $8$, and $16$, and $D=2$ receive antennas.
The transmitted symbol $x$ is binary, the channel coefficients are modeled as i.i.d. zero-mean complex Gaussian and, for simplicity, the additive disturbance is considered white.
The results that we present are averages over $1,000$ randomly generated channel realizations.
In Fig.~\ref{fig:MIMO_M4}, we plot the bit error rate (BER) of $4$PSK transmit beamforming and coherent detection after maximum-SNR filtering at the receiver, for different values of the number $N$ of transmit antennas.
In Figs.~\ref{fig:MIMO_M8} and~\ref{fig:MIMO_M16}, we repeat for $8$PSK and $16$PSK beamforming codebooks.
As expected, the BER decreases monotonically with $N$, due to increased space diversity.
We note that the optimal beamformer $\mathbf{w}_{\text{opt}}$ cannot be implemented in reasonably small time through exponential-complexity exhaustive search while the proposed algorithm computes $\mathbf{w}_{\text{opt}}$ with complexity ${\mathcal O}\left(N^4\right)$.
For example, for the case $M=8$ and $N=20$ in Fig.~\ref{fig:MIMO_M8}, the exhaustive-search computation of $\mathbf{w}_{\text{opt}}$ requires a search among $M^{N-1}\simeq1.4\times10^{17}$ vectors of length $20$ while the proposed computation of $\mathbf{w}_{\text{opt}}$ performs a search among $N+\binom{N}{3}\left(\frac{M}{2}\right)^2+N(N-1)\left(\frac{M}{2}-1\right)\simeq1.9\times10^4$ vectors of length $20$.

\section{Conclusions}

In this work, we presented a new algorithm for the computation of the $M$PSK sequence that maximizes
a fixed-rank positive semidefinite quadratic form.
Such a form is a special case of
the Rayleigh quotient of a complex matrix.
Our algorithm utilizes auxiliary continuous-valued angles and partitions the resulting continuous space of solutions into a set of regions, each of which corresponds to a distinct $M$PSK sequence.
The sequence that maximizes the
quadratic form
is shown to belong to this set of sequences which defines the new feasible set for the initial optimization problem.
Interestingly, the feasible set of candidates sequences that is identified by the proposed algorithm has cardinality polynomial in the matrix size and is constructed by the proposed algorithm with complexity polynomial in the matrix size, if the matrix rank is fixed.
Therefore, our algorithm stands as another proof-by-construction of the polynomial solvability of the fixed-rank Rayleigh quotient maximization by an $M$PSK sequence.
Moreover, the analysis of our proposed algorithm clarified its superiority to the current state of the art in terms of complexity, computational cost, memory requirement, and parallelizability.

\appendices
\section{\textbf{Proof of eq.~\eqref{eq_009}}}

To prove that~(\ref{eq_005}) is equivalent to~(\ref{eq_009}), that is,
$\max_{\mathbf{s}\in\mathcal{A}_M^N}\max_{\boldsymbol{\phi}\in\Phi^{2D-2}\times(-\pi,\pi]}\Re\left\{\mathbf{s}^{\mathcal H}\mathbf{V}\mathbf{c}(\boldsymbol{\phi})\right\}=\max_{{\bf s}\in{\mathcal A}_M^N}\max_{\boldsymbol{\phi}\in\Phi^{2D-2}\times\left(-\frac{\pi}{M},\frac{\pi}{M}\right]}\Re\left\{\mathbf{s}^\mathcal{H}\mathbf{V}\mathbf{c}(\boldsymbol{\phi})\right\},$
it suffices to show that, for any ${\bf s}\in{\mathcal A}_M^N$ and $\boldsymbol{\phi}\in\Phi^{2D-2}\times\left(-\pi,\pi\right]$, there exist $\tilde{\bf s}\in{\mathcal A}_M^N$ and $\tilde{\boldsymbol{\phi}}\in\Phi^{2D-2}\times\left(-\frac{\pi}{M},\frac{\pi}{M}\right]$, such that ${\bf s}^{\mathcal H}{\bf V}{\bf c}(\boldsymbol{\phi})=\tilde{\bf s}^{\mathcal H}{\bf V}{\bf c}(\tilde{\boldsymbol{\phi}})$.

Indeed, consider arbitrary ${\bf s}\in{\mathcal A}_M^N$ and $\boldsymbol{\phi}\in\Phi^{2D-2}\times\left(-\pi,\pi\right]$ and define
\begin{equation}
\begin{split}
&\theta\eqdef m\frac{2\pi}{M},\;\;m=0,1,\ldots,M-1,\\
&\text{such that }\\
&-\frac{\pi}{M}+m\frac{2\pi}{M}<\arg\left\{c_D(\boldsymbol\phi)\right\}\leq\frac{\pi}{M}+m\frac{2\pi}{M},
\end{split}
\label{eq:theta}
\end{equation}
where $c_D(\boldsymbol\phi)$ is the $D$th element of ${\bf c}(\boldsymbol\phi)$.
Let $\tilde{\boldsymbol{\phi}}\in\Phi^{2D-2}\times(-\pi,\pi]$ be the angle vector that consists of the spherical coordinates of ${\bf c}(\boldsymbol{\phi})e^{-j\theta}$, that is,
\begin{equation}
{\bf c}(\tilde{\boldsymbol{\phi}})\eqdef{\bf c}(\boldsymbol{\phi})e^{-j\theta},
\label{eq:ctilde}
\end{equation}
and define
\begin{equation}
\tilde{\bf s}\eqdef{\bf s}e^{-j\theta}.
\label{eq:stilde}
\end{equation}
Then,
\begin{enumerate}
\item[{\it(i)}]
due to~(\ref{eq:theta}) and~(\ref{eq:ctilde}), $\arg\{c_D(\tilde{\boldsymbol{\phi}})\}\in\left(-\frac{\pi}{M},\frac{\pi}{M}\right]\Rightarrow\tilde\phi_{2D-1}\in\left(-\frac{\pi}{M},\frac{\pi}{M}\right]$, since, due to~(\ref{eq_002}), $\arg\{c_D({\boldsymbol{\phi}})\}=\phi_{2D-1}$, for any ${\boldsymbol{\phi}}\in\Phi^{2D-2}\times(-\pi,\pi]$,
\item[{\it(ii)}]
due to~(\ref{eq:theta}) and~(\ref{eq:stilde}), $\tilde{\bf s}\in{\mathcal A}_M^N$, and
\item[{\it(iii)}]
$\tilde{\bf s}^{\mathcal H}{\bf V}{\bf c}(\tilde{\boldsymbol{\phi}})\overset{(\ref{eq:ctilde})}{\underset{(\ref{eq:stilde})}{=}}\left({\bf s}e^{-j\theta}\right)^{\mathcal H}{\bf V}{\bf c}({\boldsymbol{\phi}})e^{-j\theta}={\bf s}^{\mathcal H}{\bf V}{\bf c}({\boldsymbol{\phi}})$.
\hfill$\Box$
\end{enumerate}

\section{\textbf{Proof of Proposition~\ref{prop1}}}

{\it(i)}
Consider ${\mathcal I}=\{i_1,i_2,\ldots,i_{2D-1}\}$ and $2D-1$ hypersurfaces ${\mathcal H}(\tilde{\bf V}_{i_1,:}),{\mathcal H}(\tilde{\bf V}_{i_2,:}),\ldots,{\mathcal H}(\tilde{\bf V}_{i_{2D-1},:})$ that correspond to $2D-1$ rows of $\tilde{\bf V}$.
Since each hypersurface ${\mathcal H}(\tilde{\bf V}_{i,:})$ is described by $\tilde{\bf V}_{i,:}\tilde{\bf c}(\boldsymbol{\phi})=0$, $i=i_1,i_2,\ldots,i_{2D-1}$, their intersection(s) can be determined by solving $\tilde{\bf V}_{{\mathcal I},:}\tilde{\mathbf{c}}(\boldsymbol{\phi})={\bf 0}$.
The latter is satisfied if and only if $\tilde{\bf c}(\boldsymbol{\phi})$ belongs to the null space of $\tilde{\bf V}_{{\mathcal I},:}$ which is denoted by ${\mathcal N}(\tilde{\bf V}_{{\mathcal I},:})$ and has dimension greater than or equal to one, since $\text{rank}(\tilde{\bf V }_{{\mathcal I},:})\leq2D-1$.
Let
$\tilde{\bf V }_{{\mathcal I},:}={\bf U}{\bf\Sigma}{\bf W}^T$
be the singular value decomposition of $\tilde{\bf V}_{{\mathcal I},:}$, where ${\bf U}\in\mathbb{R}^{(2D-1)\times(2D-1)}$ and ${\bf W}\in{\mathbb R}^{2D\times2D}$ are orthonormal matrices containing the left and right, respectively, singular vectors and
${\bf\Sigma}=\left[\text{diag}\left(\sigma_1,\sigma_2,\ldots,\sigma_{2D-1}\right)\;{\bf 0}\right]_{(2D-1)\times2D}$
contains the singular values $\sigma_1\geq\sigma_2\geq\ldots\geq\sigma_{2D-1}\geq0$.
We consider two cases.
\begin{itemize}
\item
If $\sigma_{2D-1}>0$, then ${\mathcal N}(\tilde{\bf V}_{{\mathcal I},:})=\{\delta{\bf W}_{:,2D}: \; \delta\in \mathbb{R} \}$ which implies that $\tilde{\bf c}(\boldsymbol{\phi})=\frac{{\bf W}_{:,2D}}{\left\|{\bf W}_{:,2D}\right\|}$ or $-\frac{{\bf W}_{:,2D}}{\left\|{\bf W}_{:,2D}\right\|}$.
Since we require $\boldsymbol{\phi}\in\Phi^{2D-2}\times\left(-\frac{\pi}{M},\frac{\pi}{M}\right]$, only one solution $\pm\frac{{\bf W}_{:,2D}}{\left\|{\bf W}_{:,2D}\right\|}$ is valid and its spherical coordinates uniquely determine $\boldsymbol\phi$.
\item
If $\sigma_{2D-1}=0$, then $\text{rank}(\tilde{\bf V}_{{\mathcal I},:})<2D-1$ which implies that there are uncountably many solutions for $\tilde{\bf c}(\boldsymbol{\phi})$ that satisfy $\tilde{\bf V}_{{\mathcal I},:}\tilde{\bf c}(\boldsymbol{\phi})={\bf 0}$ subject to $\boldsymbol{\phi}\in\Phi^{2D-2}\times\left(-\frac{\pi}{M},\frac{\pi}{M}\right]$.
\end{itemize}

{\it(ii)}
Consider a row of ${\bf V}$, say ${\bf V}_{n,:}$, and two hypersurfaces that originate from it (i.e., they are rotated versions of it), say the ones that correspond to decision boundaries ${\mathcal B}_k^{(n)}$ and ${\mathcal B}_l^{(n)}$ where $0\leq k<l\leq\frac{M}{2}-1$.
Note that the latter inequality implies that $M\geq4$.%
\footnote{If $M=2$, i.e., ${\bf s}$ is a binary sequence, then we do not meet these degenerate cases where the intersection of hypersurfaces is not a point. This is another point that significantly differentiates the present work from the work for the binary case in~\cite{RankD:00}.}

Then, according to~(\ref{eq_011}), the intersection of the two hypersurfaces contains all points $\boldsymbol\phi\in\Phi^{2D-2}\times\left(-\frac{\pi}{M},\frac{\pi}{M}\right]$ that satisfy
\begin{align}
&\Im\left\{\begin{bmatrix}e^{-j\pi\frac{2k+1}{M}}\\e^{-j\pi\frac{2l+1}{M}}\end{bmatrix}\mathbf{V}_{n,:}\mathbf{c}(\boldsymbol{\phi})\right\}={\bf 0}\nonumber\\
\Leftrightarrow&\begin{bmatrix}\cos\left(\frac{\pi}{M}(2k+1)\right)\;\;-\sin\left(\frac{\pi}{M}(2k+1)\right)\\\cos\left(\frac{\pi}{M}(2l+1)\right)\;\;-\sin\left(\frac{\pi}{M}(2l+1)\right)\end{bmatrix}\begin{bmatrix}\Im\left\{{\bf V}_{n,:}{\bf c}(\boldsymbol\phi)\right\}\\\Re\left\{{\bf V}_{n,:}{\bf c}(\boldsymbol\phi)\right\}\end{bmatrix}\nonumber\\
&={\bf 0}.
\label{eq:cosIR}
\end{align}
The determinant of the leftmost matrix in~(\ref{eq:cosIR}) is $\sin\!\left(\frac{2\pi}{M}(k-l)\right)<0$, since $0\leq k<l\leq\frac{M}{2}-1$.
Hence,~(\ref{eq:cosIR}) is equivalent to
\begin{equation}
\begin{bmatrix}\Im\left\{{\bf V}_{n,:}{\bf c}(\boldsymbol\phi)\right\}\\\Re\left\{{\bf V}_{n,:}{\bf c}(\boldsymbol\phi)\right\}\end{bmatrix}={\bf 0}\Leftrightarrow{\bf V}_{n,:}{\bf c}(\boldsymbol\phi)=0.
\label{eq:Vc}
\end{equation}
Since the above equation is independent of the selection of $k,l$, that is, the pair of rotated versions of ${\bf V}_{n,:}$, we conclude that the intersection of any two or more rotated hypersurfaces is common and consists of all points $\boldsymbol\phi\in\Phi^{2D-2}\times\left(-\frac{\pi}{M},\frac{\pi}{M}\right]$ that satisfy~(\ref{eq:Vc}), hence, it is a $(2D-3)$-manifold in the $(2D-1)$-dimensional space.
\hfill$\Box$

\section{\textbf{Proof of Proposition~\ref{prop2}}}

{\it(i)}
Consider an angle vector $\boldsymbol\phi$ with $\phi_{2D-1}=-\frac{\pi}{M}$ and arbitrary $\boldsymbol\phi_{1:2D-2}\in\Phi^{2D-2}$ and denote by $\tilde{\boldsymbol\phi}\in\Phi^{2D-2}\times(-\pi,\pi]$ the spherical coordinates of $e^{j\frac{2\pi}{M}}{\bf c}(\boldsymbol\phi_{1:2D-2},-\frac{\pi}{M})$, that is,
\begin{equation}
{\bf c}(\tilde{\boldsymbol\phi})=e^{j\frac{2\pi}{M}}{\bf c}\left(\boldsymbol\phi_{1:2D-2},-\frac{\pi}{M}\right).
\label{eq:cphie}
\end{equation}
Then, due to~(\ref{eq_002}),
$\tilde\phi_{2D-1}=\arg\{c_D(\tilde{\boldsymbol{\phi}})\}\overset
{(\ref{eq:cphie})}{=}\frac{2\pi}{M}+\arg\{c_D(\boldsymbol\phi_{1:2D-2},-\frac{\pi}{M})\}\overset{(\ref{eq_002})}{=}\frac{2\pi}{M}-\frac{\pi}{M}=\frac{\pi}{M},$
that is,
\begin{equation}
{\bf c}\left(\tilde{\boldsymbol\phi}_{1:2D-2},\frac{\pi}{M}\right)=e^{j\frac{2\pi}{M}}{\bf c}\left(\boldsymbol\phi_{1:2D-2},-\frac{\pi}{M}\right).
\label{eq:cphipiM}
\end{equation}
Moreover,
${\bf d}({\bf V};\boldsymbol\phi_{1:2D-2},-\frac{\pi}{M})=\argmax_{\mathbf{s}\in\mathcal{A}^N_M}\Re\{\mathbf{s}^\mathcal{H}\mathbf{V}$ $\mathbf{c}(\boldsymbol{\phi}_{1:2D-2},-\frac{\pi}{M})\}={\displaystyle\argmax_{\mathbf{s}\in\mathcal{A}^N_M}}\Re\{({\bf s}e^{j\frac{2\pi}{M}})^{\mathcal H}{\bf V}e^{j\frac{2\pi}{M}}{\bf c}(\boldsymbol{\phi}_{1:2D-2},$
$-\frac{\pi}{M})\}\overset{(\ref{eq:cphipiM})}{=}e^{-j\frac{2\pi}{M}}{\displaystyle\argmax_{\mathbf{s}\in\mathcal{A}^N_M}}\Re\{{\bf s}^{\mathcal H}{\bf V}{\bf c}(\tilde{\boldsymbol{\phi}}_{1:2D-2},\frac{\pi}{M})\}=e^{-j\frac{2\pi}{M}}$
${\bf d}\left({\bf V};\tilde{\boldsymbol\phi}_{1:2D-2},\frac{\pi}{M}\right).$

{\it(ii)}
Consider an angle vector $\boldsymbol\phi$ with $\phi_{2D-2}=\frac{\pi}{2}$ and arbitrary $\boldsymbol\phi_{1:2D-3}\in\Phi^{2D-3}$ and $\phi_{2D-1}\in\left(-\frac{\pi}{M},\frac{\pi}{M}\right]$.
Then,
\begin{equation}
{\bf c}\left({\boldsymbol\phi}_{1:2D-3},\frac{\pi}{2},\phi_{2D-1}\right)=\begin{bmatrix}
{\bf c}\left({\boldsymbol\phi}_{1:2D-3}\right)\\
0
\end{bmatrix}
\label{eq:cphi0}
\end{equation}
and
${\bf d}({\bf V};{\boldsymbol\phi}_{1:2D-3},\frac{\pi}{2},\phi_{2D-1})={\displaystyle\argmax_{\mathbf{s}\in\mathcal{A}^N_M}}\Re\{\mathbf{s}^\mathcal{H}\mathbf{V}\mathbf{c}({\boldsymbol\phi}_{1:2D-3},$
$\frac{\pi}{2},\phi_{2D-1})\}\overset{(\ref{eq:cphi0})}{=}\argmax_{\mathbf{s}\in\mathcal{A}^N_M}\Re\left\{\mathbf{s}^\mathcal{H}\mathbf{V}\begin{bmatrix}
{\bf c}({\boldsymbol\phi}_{1:2D-3})\\
0
\end{bmatrix}\right\}=\argmax_{\mathbf{s}\in\mathcal{A}^N_M}\Re\{{\bf s}^{\mathcal H}{\bf V}_{:,1:D-1}{\bf c}({\boldsymbol\phi}_{1:2D-3})\}={\bf d}({\bf V}_{:,1:D-1};$ ${\boldsymbol\phi}_{1:2D-3}).$

{\it(iii)}
Consider an angle vector $\boldsymbol\phi$ with $\phi_{2D-2}=-\frac{\pi}{2}$ and arbitrary $\boldsymbol\phi_{1:2D-3}\in\Phi^{2D-3}$ and $\phi_{2D-1}\in\left(-\frac{\pi}{M},\frac{\pi}{M}\right]$.
Then,
\begin{align}
{\bf c}\left({\boldsymbol\phi}_{1:2D-3},-\frac{\pi}{2},\phi_{2D-1}\right)&=-\begin{bmatrix}
{\bf c}\left(-{\boldsymbol\phi}_{1:2D-3}\right)\\
0
\end{bmatrix}\label{eq:cphihat}\\
&\overset{(\ref{eq:cphi0})}{=}-{\bf c}\left(-{\boldsymbol\phi}_{1:2D-3},\frac{\pi}{2},\widehat\phi_{2D-1}\right)\nonumber
\end{align}
for any $\widehat\phi_{2D-1}\in(-\frac{\pi}{M},\frac{\pi}{M}]$.
Hence,
${\bf d}({\bf V};{\boldsymbol\phi}_{1:2D-3},-\frac{\pi}{2},$ $\phi_{2D-1})=\argmax_{\mathbf{s}\in\mathcal{A}^N_M}\Re\{\mathbf{s}^\mathcal{H}\mathbf{V}\mathbf{c}({\boldsymbol\phi}_{1:2D-3},-\frac{\pi}{2},\phi_{2D-1})\}$
$\overset{(\ref{eq:cphihat})}{=}\argmax_{\mathbf{s}\in\mathcal{A}^N_M}\Re\{(-{\bf s})^\mathcal{H}\mathbf{V}\mathbf{c}(-{\boldsymbol\phi}_{1:2D-3},\frac{\pi}{2},\widehat\phi_{2D-1})\}=$ $-\argmax_{\mathbf{s}\in\mathcal{A}^N_M}\Re\{{\bf s}^{\mathcal H}\mathbf{V}$ $\mathbf{c}(-{\boldsymbol\phi}_{1:2D-3},\frac{\pi}{2},\widehat\phi_{2D-1})\}=-{\bf d}({\bf V};$ $-{\boldsymbol\phi}_{1:2D-3},\frac{\pi}{2},\widehat\phi_{2D-1}).$

{\it(iv)}
Consider an angle vector $\boldsymbol\phi$ with $\phi_{2D-2}=\pm\frac{\pi}{2}$ and arbitrary $\boldsymbol\phi_{1:2D-3}\in\Phi^{2D-3}$ and $\phi_{2D-1}\in(-\frac{\pi}{M},\frac{\pi}{M}]$.
Then,
\begin{equation}
{\bf c}\left({\boldsymbol\phi}_{1:2D-3},\pm\frac{\pi}{2},\phi_{2D-1}\right)={\bf c}\left({\boldsymbol\phi}_{1:2D-3},\pm\frac{\pi}{2},\widehat\phi_{2D-1}\right)
\label{eq:cphihat2}
\end{equation}
for any $\widehat\phi_{2D-1}\in(-\frac{\pi}{M},\frac{\pi}{M}]$, since $\cos(\pm\frac{\pi}{2})=0$.
Hence,
${\bf d}({\bf V};$ ${\boldsymbol\phi}_{1:2D-3},\pm\frac{\pi}{2},\phi_{2D-1})=\argmax_{\mathbf{s}\in\mathcal{A}^N_M}\Re\{\mathbf{s}^\mathcal{H}\mathbf{V}\mathbf{c}({\boldsymbol\phi}_{1:2D-3},$
$\pm\frac{\pi}{2},\phi_{2D-1})\}\overset{(\ref{eq:cphihat2})}{=}\argmax_{\mathbf{s}\in\mathcal{A}^N_M}\Re\{{\bf s}^{\mathcal H}\mathbf{V}\mathbf{c}({\boldsymbol\phi}_{1:2D-3},\pm\frac{\pi}{2},$
$\widehat\phi_{2D-1})\}={\bf d}({\bf V};{\boldsymbol\phi}_{1:2D-3},\pm\frac{\pi}{2},\widehat\phi_{2D-1}).$
\hfill$\Box$

\section{\textbf{Proof of Proposition~\ref{prop3}}}

If $D=N$, then the cardinality expression in~(\ref{eq_prop4_00}) becomes
\begin{align}
&|\mathcal{S}(\mathbf{V})|=\\
&\sum_{d = 1}^{N} \sum_{i = 0}^{d-1} \binom{N}{i}\binom{N - i}{2(d-i) - 1}\left(\frac{M}{2}\right)^{2(d-i)-2}\left(\frac{M}{2} - 1\right)^{i}.\nonumber
\end{align}
By interchanging summations and making some variable substitutions, the above equation is transformed into the equivalent form%
\footnote{As in most binomial and multinomial proofs, quantities of the form $ 0^0 $ are assumed to be equal to 1.}
\begin{align}
&|\mathcal{S}(\mathbf{V})| = \sum_{i = 0}^{N} \sum_{\substack{d = 0, \\ d \;\rm{odd}}}^{N-i} \binom{N}{i}\binom{N - i}{d}\left(\frac{M}{2}\right)^{d-1}\left(\frac{M}{2} - 1\right)^i\nonumber\\
&= \left(\frac{M}{2}\right)^{-1}\sum_{i = 0}^{N} \sum_{\substack{d = 0, \\ d \;\rm{odd}}}^{N-i} \binom{N}{i}\binom{N - i}{d}\left(\frac{M}{2}\right)^{d}\left(\frac{M}{2} - 1\right)^i\nonumber\\
&= \left(\frac{M}{2}\right)^{-1}\sum_{i = 0}^{N} \binom{N}{i}\left(\frac{M}{2} - 1\right)^i\sum_{\substack{d = 0, \\ d \;\rm{odd}}}^{N-i} \binom{N - i}{d}\left(\frac{M}{2}\right)^{d}.
\label{eq_prop4_01}
\end{align}
We consider the binomial formula
\begin{align}
(a + b)^n = \sum_{k=0}^n \binom{n}{k} a^k b^{n-k}
\label{eq_prop4_02}
\end{align}
which holds for any $a,b\in{\mathbb R}$ and $n,k\in{\mathbb N}$, and the fact that the sum of the coefficients of the odd terms of the expansion of $(a+b)^n$ is equal to the sum of the coefficients of the even terms.
Then,
\begin{align}
\sum_{\substack{d = 0, \\ d \;\rm{odd}}}^{N-i} \binom{N - i}{d}\left(\frac{M}{2}\right)^{d} = \frac{1}{2} \sum_{\substack{d = 0}}^{N-i} \binom{N - i}{d}\left(\frac{M}{2}\right)^{d}.
\end{align}
Therefore,~(\ref{eq_prop4_01}) becomes
\begin{equation}
\begin{split}
&|\mathcal{S}(\mathbf{V}_{N \times N})|\\
&= \left(\frac{M}{2}\right)^{-1}\sum_{i = 0}^{N} \binom{N}{i}\left(\frac{M}{2} - 1\right)^i\sum_{\substack{d = 0, \\ d \;\rm{odd}}}^{N-i} \binom{N - i}{d}\left(\frac{M}{2}\right)^{d}\\
&= \left(\frac{M}{2}\right)^{-1}\sum_{i = 0}^{N} \binom{N}{i}\left(\frac{M}{2} - 1\right)^i\frac{1}{2} \sum_{\substack{d = 0}}^{N-i} \binom{N - i}{d}\left(\frac{M}{2}\right)^{d}\\
&= (M)^{-1}\sum_{i = 0}^{N} \binom{N}{i}\left(\frac{M}{2} - 1\right)^i \sum_{\substack{d = 0}}^{N-i} \binom{N - i}{d}\left(\frac{M}{2}\right)^{d}\\
&\overset{(\ref{eq_prop4_02})}{=}(M)^{-1}\sum_{i = 0}^{N} \binom{N}{i}\left(\frac{M}{2} - 1\right)^i \left(\frac{M}{2}+1\right)^{N-i}\\
&= (M)^{-1}\left(\frac{M}{2} - 1 + \frac{M}{2} + 1\right)^N= (M)^{-1}M^N = M^{N-1}.
\end{split}
\end{equation}
\hfill$\Box$

\newpage

\begin{IEEEbiography}{Anastasios Kyrillidis} (S'11) received the Diploma (five-year program) and M.Sc. degrees in electronic and computer engineering from the Technical University of Crete, Chania, Greece, in 2008 and 2010, respectively.
Currently, he is a graduate student at Ecole Polytechnique Federale de Lausanne.
His research interests include machine learning and high-dimensional data analysis and mining.
\end{IEEEbiography}

\begin{IEEEbiography}{George N. Karystinos} (S'98-M'03) was born in Athens, Greece, on April 12, 1974.
He received the Diploma degree in computer science and engineering (five-year program) from the University of Patras, Patras, Greece, in 1997 and the Ph.D. degree in electrical engineering from the State University of New York at Buffalo, Amherst, NY, in 2003.

From 2003 to 2005, he held an Assistant Professor position in the Department of Electrical Engineering, Wright State University, Dayton, OH.
Since 2005, he has been with the Department of Electronic and Computer Engineering, Technical University of Crete, Chania, Greece, where he is presently an Associate Professor.
His research interests are in the general areas of $L_1$-norm principal component analysis of data and signals, communication theory, and adaptive signal processing with applications to signal waveform design, low-complexity sequence detection, and secure wireless communications.

Dr. Karystinos is a member of the IEEE Signal Processing, Communications, Information Theory, and Computational Intelligence Societies.
For articles that he coauthored with students and colleagues, he received a 2001 IEEE International Conference on Telecommunications best paper award, the 2003 IEEE \textsc{Transactions On Neural Networks} Outstanding Paper Award, the 2011 IEEE International Conference on RFID-Technologies and Applications Second Best Student Paper Award, and the 2013 International Symposium on Wireless Communication Systems Best Paper Award in Signal Processing and Physical Layer Communications.
\end{IEEEbiography}

\vfill


\begin{thebibliography}{99}

\bibitem{Heath1998}
R. W. Heath, Jr. and A. Paulraj,
``A simple scheme for transmit diversity using partial channel feedback,''
in {\it Proc. IEEE Asilomar Conf. on Signals, Systems, and Comp.},
Pacific Grove, CA,
Nov. 1998,
vol. 2,
pp. 1073-1078.

\bibitem{EGT:01}
D. J. Love, R. W. Heath, Jr., and T. Strohmer,
``Grassmannian beamforming for multiple-input multiple-output wireless systems,''
{\it IEEE Trans. Inf. Theory},
vol. 49,
pp. 2735-2747,
Oct. 2003.

\bibitem{Zheng2007}
X. Zheng, Y. Xie, J. Li, and P. Stoica,
``MIMO transmit beamforming under uniform elemental power constraint,''
{\it IEEE Trans. Signal Process.},
vol. 55,
pp. 5395-5406,
Nov. 2007.

\bibitem{Code:00}
D. J. Ryan, I.V. L. Clarkson, I. B. Collings, D. Guo, and M. L. Honig,
``QAM and PSK codebooks for limited feedback MIMO beamforming,''
{\it IEEE Trans. Commun.},
vol. 57,
pp. 1184-1196,
Apr. 2009.

\bibitem{Xu2009}
M. Xu, D. Guo, and M. L. Honig,
``MIMO precoding with limited rate feedback: Simple quantizers work well,''
in {\it Proc. IEEE GLOBECOM 2009},
Honolulu, HI,
Dec. 2009.

\bibitem{EGT:02}
Y. G. Kim and N. C. Beaulieu,
``On MIMO beamforming systems using quantized feedback,''
{\it IEEE Trans. Commun.},
vol. 58,
pp. 820-827,
Mar. 2010.

\bibitem{Iran}
K.-K. Leung, C. W. Sung, M. Khabbazian, and M. A. Safari,
``Optimal phase control for equal-gain transmission in MIMO systems with scalar quantization: Complexity and algorithms,''
{\it IEEE Trans. Inf. Theory},
vol. 56,
pp. 3343-3355,
July 2010.

\bibitem{Santipach2011}
W. Santipach and K. Mamat,
``Tree-structured random vector quantization for limited-feedback wireless channels,''
{\it IEEE Trans. Wireless Commun.},
vol. 10,
pp. 3012-3019,
Sept. 2011.

\bibitem{MLSD:00}
I. Motedayen, A. Krishnamoorthy, and A. Anastasopoulos,
``Optimal joint detection/estimation in fading channels with polynomial complexity,''
{\it IEEE Trans. Inf. Theory},
vol. 53,
pp. 209-223,
Jan. 2007.

\bibitem{SIMO:00}
D. S. Papailiopoulos and G. N. Karystinos,
``Polynomial-complexity maximum-likelihood block noncoherent MPSK detection,''
in {\it Proc. IEEE ICASSP 2008},
Las Vegas, NV,
Apr. 2008,
pp. 2681-2684.

\bibitem{Xu2008}
W. Xu, M. Stojnic, and B. Hassibi,
``Low-complexity blind maximum-likelihood detection for SIMO systems with general constellations,''
in {\it Proc. IEEE ICASSP 2008},
Las Vegas, NV,
Apr. 2008,
pp. 2817-2820.

\bibitem{CG:00}
V. Pauli, L. Lampe, R. Schober, and K. Fukuda,
``Multiple-symbol differential detection based on combinatorial geometry,''
{\it IEEE Trans. Commun.},
vol. 56,
pp. 1596-1600,
Oct. 2008.

\bibitem{Wu2009}
M. Wu and P. Y. Kam,
``Sequence detection on fading channels without explicit channel estimation,''
in {\it Proc. IEEE Wireless VITAE 2009},
Aalborg, Denmark,
May 2009,
pp. 370-374.

\bibitem{Wu2010}
M. Wu and P. Y. Kam,
``Performance analysis and computational complexity comparison of sequence detection receivers with no explicit channel estimation,''
{\it IEEE Trans. Vehic. Tech.},
vol. 59,
pp. 2625-2631,
June 2010.

\bibitem{STBC}
D. S. Papailiopoulos and G. N. Karystinos,
``Maximum-likelihood noncoherent OSTBC detection with polynomial complexity,''
{\em IEEE Trans. Wireless Commun.},
vol. 9,
pp. 1935-1945,
June 2010.

\bibitem{RankD:00}
G. N. Karystinos and A. P. Liavas,
``Efficient computation of the binary vector that maximizes a rank-deficient quadratic form,''
{\it IEEE Trans. Inf. Theory},
vol. 56,
pp. 3581-3593,
July 2010.

\bibitem{CDMA:01}
G. Manglani and A. K. Chaturvedi,
``Application of computational geometry to multiuser detection in CDMA,''
{\it IEEE Trans. Commun.},
vol. 54,
pp. 204-207,
Feb. 2006.

\bibitem{Zon:00}
K. Allemand, K. Fukuda, T. M. Liebling, and E. Steiner,
``A polynomial case of unconstrained zero-one quadratic optimization,''
{\it Mathematical Programming},
vol. A-91,
pp. 49-52,
Oct. 2001.

\bibitem{Inc:00}
H. Edelsbrunner, J. O'Rourke, and R. Seidel,
``Constructing arrangements of lines and hyperplanes with applications,''
{\it SIAM J. Comput.},
vol. 15,
pp. 341-363,
May 1986.

\bibitem{Edelsbrunner-book}
H. Edelsbrunner,
{\it Algorithms in combinatorial geometry}.
New York:
Springer-Verlag,
1987.

\bibitem{Rev:00}
D. Avis and K. Fukuda,
``Reverse search for enumeration,''
{\it Discrete Applied Mathematics},
vol. 65,
pp. 21-46,
Mar. 1996.

\bibitem{Zon:01}
J.-A. Ferrez, K. Fukuda, and T. M. Liebling,
``Solving the fixed rank convex quadratic maximization in binary variables by a parallel zonotope construction algorithm,''
{\it European Journal of Operational Research},
vol. 166,
pp. 35-50,
2005.

\bibitem{Megiddo}
N. Megiddo,
``Linear programming in linear time when the dimension is fixed,''
{\it J. ACM},
vol. 31,
pp. 114-127,
Jan. 1984.

\bibitem{Mack:00}
K. M. Mackenthun, Jr.,
``A fast algorithm for multiple-symbol differential detection of MPSK,''
{\it IEEE Trans. Commun.},
vol. 42,
pp. 1471-1474,
Feb./Mar./Apr. 1994.

\bibitem{rank2}
G. N. Karystinos and D. A. Pados,
``Rank-2-optimal adaptive design of binary spreading codes,''
{\it IEEE Trans. Inf. Theory},
vol. 53,
pp. 3075-3080,
Sept. 2007.

\bibitem{rank3}
G. N. Karystinos and A. P. Liavas,
``Efficient computation of the binary vector that maximizes a rank-3 quadratic form,''
in {\it Proc. 2006 Allerton Conf. Commun., Control, and Computing},
Allerton House, Monticello, IL,
Sept. 2006,
pp. 1286-1291.

\bibitem{sweldens}
W. Sweldens,
``Fast block noncoherent decoding,''
{\it IEEE Comm. Letters},
vol. 5,
pp. 132-134,
Apr. 2001.

\end{thebibliography}
\end{document}